%% file: main.tex
\documentclass[final,3p,times,twocolumn]{elsarticle}
\pdfoutput=1 
\DeclareGraphicsExtensions{.pdf}

\usepackage{graphicx}

\usepackage{bm}% bold math

\usepackage{atlasphysics} 
\usepackage[hyperindex,breaklinks]{hyperref}

\usepackage{multirow}
\usepackage{amsmath}

\usepackage{float}

\usepackage{subfigure}

\usepackage{rotating}

\usepackage{wasysym}
\usepackage{xspace}

\input{HWWlvlvPaper_Defs.tex}

\newcommand{\atlasnote}[1]{\def\@atlasnote{#1}}

\newcommand{\mgg}{\ensuremath{m_{\gamma\gamma}}}

\newcommand{\mm}{\ensuremath{\,\mbox{mm}}}

\newcommand{\bb}{\ensuremath{b\bar{b}}}
\newcommand{\ggF}{\ensuremath{gg\to H}}

\newcommand{\ttH}{\ensuremath{gg/q\bar{q} \to t\bar{t}H}}

\newcommand{\Zjets}{$Z$+jets}
\newcommand{\mll}{\ensuremath{m_{\ell\ell}}}

\newcommand{\DeltaR}{\ensuremath{\Delta R=\sqrt{\Delta\phi^2 + \Delta\eta^2}}}
\newcommand{\ptt}{\ensuremath{p_{\mathrm{Tt}}}}

\newcommand{\hgg}{\ensuremath{H\rightarrow \gamma\gamma}} 
\newcommand{\hww}{\ensuremath{H \rightarrow WW^{*}}}

\newcommand{\hWWlnln}{\ensuremath{H{\rightarrow\,}WW^{*}{\rightarrow\,}\ell\nu\ell\nu}}
\newcommand{\hWWlvlv}{\ensuremath{H{\rightarrow\,}WW^{*}{\rightarrow\,}\ell\nu\ell\nu}}
\newcommand{\hwwlnln}{\ensuremath{H{\rightarrow\,}WW^{*}{\rightarrow\,}\ell\nu\ell\nu}}

\newcommand{\hzz}{\ensuremath{H \rightarrow ZZ^{*}}}

\newcommand{\hZZllll}{\ensuremath{H{\rightarrow\,}ZZ^{*}{\rightarrow\,}4\ell}}

\newcommand{\hzzllll}{\ensuremath{H \rightarrow ZZ^{*}\rightarrow 4\ell}}

\newcommand{\hbb}{\ensuremath{H \rightarrow \bb}}

\newcommand{\mh}{\ensuremath{m_{H}}}

\newcommand{\Wjets}{$W$+jets}

\newcommand{\dphiggjj}{\ensuremath{\Delta\phi_{\gamma\gamma,jj}}} 

\def\vMPT{\ensuremath{{\bf p}_{\rm T}^{\rm miss}}}
\def\met{\ensuremath{E_{\mathrm{T}}^{\mathrm{miss}}}}

\newcommand{\PH}{\ensuremath{H}} 
\newcommand{\PW}{\ensuremath{W}} 
\newcommand{\PZ}{\ensuremath{Z}}
\newcommand{\Pg}{\ensuremath{g}}
\newcommand{\PGg}{\ensuremath{\gamma}}
\newcommand{\PQt}{\ensuremath{t}}
\newcommand{\PAQt}{\ensuremath{\bar{t}}}
\newcommand{\PQb}{\ensuremath{b}}
\newcommand{\PAQb}{\ensuremath{\bar{b}}}
\newcommand{\PQc}{\ensuremath{c}}
\newcommand{\PAQc}{\ensuremath{\bar{c}}}
\newcommand{\PQs}{\ensuremath{s}}
\newcommand{\PAQs}{\ensuremath{\bar{s}}}
\newcommand{\PGt}{\ensuremath{\tau}}
\newcommand{\PGtm}{\ensuremath{\tau^-}}
\newcommand{\PGtp}{\ensuremath{\tau^+}}
\newcommand{\PGmm}{\ensuremath{\mu^-}}
\newcommand{\PGmp}{\ensuremath{\mu^+}}
\newcommand{\Cc}{\ensuremath{\kappa}} 
\newcommand{\Rr}{\ensuremath{\lambda}}

\newcommand{\MyggH}{\ensuremath{\Pg\Pg\PH}}

\mathchardef\mhyphen="2D

\def\vec#1{{\mbox{$\boldsymbol{#1}$}}}
\renewcommand{\ttbar}{\ensuremath{t\overline{t}}}


\usepackage{amssymb}

\usepackage{chngpage}

\biboptions{numbers,square,comma,sort&compress}

\usepackage{hypernat}

\journal{Physics Letters B}

\usepackage{preprintcover}  

\PreprintCoverPaperTitle{Measurements of Higgs boson production and couplings in diboson
 final states with the ATLAS detector at the LHC}  % paper title

\PreprintIdNumber{CERN-PH-EP-2013-103}  % CERN preprint number

\PreprintCoverAbstract{
   Measurements are presented of production properties and couplings of the recently discovered
   Higgs boson using the decays into boson pairs, \hgg, \hZZllll\ and \hWWlnln.
   The results are based on the complete $pp$ collision data sample recorded 
   by the ATLAS experiment at the CERN Large Hadron Collider at centre-of-mass energies
   of $\sqrt{s}$=7 TeV and $\sqrt{s}$=8~TeV, corresponding to an integrated luminosity of about 25\,\ifb.  
   Evidence for Higgs boson production through vector-boson fusion is reported.
    Results of combined fits probing Higgs boson couplings to fermions and bosons, 
    as well as anomalous contributions to loop-induced production and decay modes,
    are presented. All measurements are consistent with expectations 
    for the Standard Model Higgs boson.}  % paper abstract

\PreprintJournalName{Physics Letters B}  

\begin{document}

\begin{frontmatter}

\title{Measurements of Higgs boson production and couplings in diboson final states with the ATLAS detector at the LHC} 

\begin{abstract}

   Measurements are presented of production properties and couplings of the recently discovered
   Higgs boson using the decays into boson pairs, \hgg, \hZZllll\ and \hWWlnln.
   The results are based on the complete $pp$ collision data sample recorded 
   by the ATLAS experiment at the CERN Large Hadron Collider at centre-of-mass energies
   of $\sqrt{s}$=7 TeV and $\sqrt{s}$=8~TeV, corresponding to an integrated luminosity of about 25\,\ifb.  
   Evidence for Higgs boson production through vector-boson fusion is reported.
    Results of combined fits probing Higgs boson couplings to fermions and bosons, 
    as well as anomalous contributions to loop-induced production and decay modes,
    are presented. All measurements are consistent with expectations 
    for the Standard Model Higgs boson.

\end{abstract}

\end{frontmatter}

\hyphenation{ATLAS}

\section{Introduction\label{sec:introduction}}
\input{Introduction}

\section {Data sample and event reconstruction\label{sec:physobj}}
\input{PhysObj}

\section{Signal and background simulation\label{sec:MC}}
\input{Samples}

\section{\texorpdfstring{The $\hgg$ channel}{H->gg}\label{sec:hgg}}
\input{Hggcoupl2013.tex}
\section{\texorpdfstring{The \hZZllll\ channel}{H->4l}\label{sec:hzz}}
\input{H4lcoupl2013.tex}

\section{\texorpdfstring{The \hWWlvlv\ channel}{H->lvlv}\label{sec:hww}}
\input{Hwwcoupl2013.tex}

\section {Higgs boson property measurements\label{sec:results}}
The results from the individual channels described in the previous sections
are combined here to extract information about the Higgs boson
mass, production properties and couplings. 

\subsection{Statistical method\label{stat}}
\input{Methods.tex}

\subsection{Mass and production strength\label{mass-mu}}
\input{Mass-mu.tex}

\subsection{Evidence for production via vector-boson fusion\label{VBF}}
\input{VBF.tex}

\subsection{Couplings measurements\label{Couplings}}
\input{CouplingIntro.tex}

\subsubsection{Couplings to fermions and bosons\label{CVCF}}
\input{CVCF.tex}

\subsubsection{Ratio of couplings to the $W$ and $Z$ bosons\label{Custodial}}
\input{Custodial.tex}

\subsubsection{Constraints on production and decay loops\label{kgkg}}
\input{KgammaKg.tex}

\subsubsection{Summary\label{sum}}
\input{Results.tex}

\section{Conclusions\label{sec:conclusions}}

   Data recorded by the ATLAS experiment at the CERN Large Hadron Collider
   in 2011 and 2012, corresponding to an integrated
  luminosity of up to 25\,\ifb\, at $\sqrt{s}=7\TeV$ and $\sqrt{s}=8\TeV$, have been 
  analysed to determine 
  several properties of the recently discovered Higgs boson using the
  $H\to\gamma\gamma$, \hZZllll\ and \hWWlnln\ decay modes.  The reported results 
  include measurements of the mass and signal strength, evidence for production through vector-boson
  fusion, and constraints on couplings to bosons and fermions as well as on anomalous contributions 
  to loop-induced processes. The precision exceeds previously published results in several cases. 
  All measurements are consistent with expectations for the Standard Model Higgs boson.

\input{acknowledgements.tex}

\bibliographystyle{atlasBibStyleWithTitle} 

\bibliography{main}

\clearpage 
\onecolumn 
\input{atlas_authlist} 
\clearpage 

\end{document}

%% file: HWWlvlvPaper_Defs.tex
\newcommand{\metrel}{\ensuremath{E_{\rm T,\ rel}^{\rm miss}}}
\newcommand{\mT}{\ensuremath{m_{\rm T}}}
\newcommand{\pTll}{\ensuremath{p_{\rm T}^{\ell\ell}}}

\newcommand{\ptll}{\pTll}
\newcommand{\vpTll}{\ensuremath{{\bf p}_{\rm T}^{\ell\ell}}}
\newcommand{\vpTtot}{\ensuremath{{\bf p}_{\rm T}^{\rm tot}}\xspace}

\newcommand{\pTvec}{\ensuremath{{\bf p}_{\rm T}}}
\newcommand{\DPhill}{\ensuremath{\Delta\phi_{\ell\ell}}\xspace}
\newcommand{\dphillMET}{\ensuremath{|\,\Delta\phi_{\ell\ell,\ E_{\mathrm{T}}^{\mathrm{miss}}}\,|}\xspace}

\newcommand{\pTtot}{\ensuremath{{\bf p}_{\rm T}^{\rm tot}}\xspace}

\newcommand{\Dyjj}{\ensuremath{|\,\Delta{y}_{jj}\,|}}

\newcommand{\emme}{\ensuremath{e\mu}}
\newcommand{\eemm}{\ensuremath{ee{/}\mu\mu}}
\newcommand{\Njet}{\ensuremath{N_\textrm{jet}}}

\newcommand{\ZeroJet}{\ensuremath{N_\textrm{jet}\,{=}\,0}}
\newcommand{\OneJet}{\ensuremath{N_\textrm{jet}\,{=}\,1}}
\newcommand{\TwoJet}{\ensuremath{N_\textrm{jet}\,{\ge}\,2}}

\newcommand{\ZeroOneJetSimple}{\ensuremath{N_\textrm{jet}\,{\le}\,1}}
\newcommand{\OneTwoJetSimple}{\ensuremath{N_\textrm{jet}\,{\ge}\,1}}

\newcommand{\AllJet}{\ZeroJet, ${=}\,1$, and ${\ge}\,2$}

\newcommand{\ZDY}{\ensuremath{Z/\gamma^\ast}}

\newcommand{\vMET}{\ensuremath{{\bf E}_{\rm T}^{\rm miss}}}
\newcommand{\Wg}{\ensuremath{W\gamma^{(\ast)}}}

\def\MPTRel{\ensuremath{p_{\mathrm{T,\ rel}}^{\mathrm{miss}}}}
\newcommand{\frecoil}{\ensuremath{f_{\rm recoil}}\xspace}

\newcommand{\METstvf}{\ensuremath{{E}_{\rm T, \ STVF}^{\rm miss}}}

\def\dbline{\noalign{\vskip 0.10truecm\hrule\vskip 0.05truecm\hrule\vskip 0.10truecm}}
\def\sgline{\noalign{\vskip 0.10truecm\hrule\vskip 0.10truecm}}

\def\NompzerosigmaExpComb{\ensuremath{3.8}}	% mH=125.5

\def\MinpzerosigmaComb{\ensuremath{4.1}}
\def\MinpzeromassComb{\ensuremath{140}}

%% file: Introduction.tex
The discovery of a new particle of mass about 125~GeV in the search for the Standard Model
(SM) Higgs boson at the CERN Large Hadron Collider (LHC)~\cite{1748-0221-3-08-S08001}, 
reported in July 2012 by the
ATLAS~\cite{paper2012ichep} and CMS~\cite{CMSpaper} Collaborations,
is a milestone in the quest to understand the origin of electroweak symmetry
breaking~\cite{Englert:1964et,Higgs:1964ia,Higgs:1964pj,Guralnik:1964eu,Higgs:1966ev,Kibble:1967sv}. 

 This paper presents measurements of several properties of the 
 newly observed particle, including its mass,
 production strengths and couplings to fermions and bosons, using
 diboson final states:\footnote{Throughout this paper, the symbol $\ell$ stands for
 electron or muon.}~\hgg, \hZZllll, and \hWWlnln.
 Spin studies are reported elsewhere~\cite{ATLASspinpaper}. 
  Due to the outstanding performance of the LHC accelerator 
throughout 2012, the present data sample 
is a factor of $\sim$2.5 larger than that used in
 Ref.~\cite{paper2012ichep}. With these additional 
data, many aspects of the ATLAS studies have been improved: 
 several experimental uncertainties have been reduced and new
 exclusive analyses have been included. In particular, event 
 categories targeting specific production modes have been introduced, 
 providing enhanced sensitivity to different Higgs boson couplings.   
 
  The results reported here are based 
 on the data samples recorded with the ATLAS detector~\cite{atlas-det}
 in 2011 (at $\sqrt{s}=7\TeV$) and 2012 (at  $\sqrt{s}=8\TeV$), corresponding to 
 integrated luminosities of about 4.7\,\ifb\ and 20.7\,\ifb, respectively.  
  Similar studies, including also fermionic decays, 
 have been reported recently by the CMS Collaboration using 
 a smaller dataset~\cite{CMSpapernew}. 
 
   This paper is organised as follows. Section~\ref{sec:physobj} describes the 
data sample and the event reconstruction. 
 Section~\ref{sec:MC} summarises 
 the Monte Carlo (MC) samples used to model signal and
background processes.  The analyses of the three decay channels
are presented in Sections~\ref{sec:hgg}--\ref{sec:hww}. 
 Measurements of the Higgs boson mass, production properties and couplings
are discussed in Section~\ref{sec:results}.
 Section~\ref{sec:conclusions} is devoted to the conclusions.

%% file: PhysObj.tex
   After data quality requirements, the integrated luminosities of the
   samples used for the studies reported here are about 4.7\,\ifb\ 
  in 2011 and 20.7\,\ifb\ in 2012, with uncertainties given in Table~\ref{tab:commonsys}
  (determined as described in Ref.~\cite{lumi2011}).  
   Because of the high LHC peak luminosity (up to 
  $7.7\times10^{33}$~cm$^{-2}$~s$^{-1}$ in 2012) and the 50~ns bunch spacing, the 
   number of proton--proton interactions occurring 
  in the same bunch crossing is large (on average 20.7, up to about 40).
  This ``pile-up" of events
  requires the use of dedicated algorithms and corrections to mitigate 
  its impact on the reconstruction of {\it e.g.} leptons, photons and jets. 
 
 \begin{table}[!htbp]
   \centering
   \caption{Main sources of experimental uncertainty, and of theoretical 
    uncertainty on the signal yield, common to the three 
     channels considered in this study.
     Theoretical uncertainties are given for a SM Higgs boson of mass
     $m_H=125$~GeV and are taken from 
     Refs.~\cite{YellowReport,LHCHiggsCrossSectionWorkingGroup:2012vm,HiggsXSWGWebPage}.
     ``QCD scale" indicates (here and throughout this paper) 
     QCD renormalisation and factorisation scales and ``PDFs" indicates parton distribution functions. 
     The ranges for the experimental uncertainties cover the variations 
     with \pt\ and  $\eta$.}
   \vspace{0.3cm}
   \scalebox{0.9}{
   \begin{tabular}{ll}
     \hline\hline
     Source (experimental)  & Uncertainty (\%)  \\
     \quad Luminosity  &  $\pm$1.8 (2011), $\pm$3.6 (2012) \\     
     \quad Electron efficiency & $\pm$2--5 \\
     \quad Jet energy scale & $\pm$1--5 \\
     \quad Jet energy resolution  & $\pm$2--40 \\
%     $b$-jet tagging efficiency & $\pm$5--12 \\  
     \hline
     Source (theory)  & Uncertainty (\%) \\
     \quad QCD scale  &  $\pm8$ (ggF), $\pm1$(VBF, VH), $^{+4}_{-9}$ (ttH)\\
     \quad PDFs + $\alpha_{s}$    &  $\pm$8  (ggF, ttH), $\pm 4$ (VBF, VH)\\ 
     \hline\hline \\
   \end{tabular}
   } \label{tab:commonsys}
 \end{table}  
   For the \hZZllll\ and \hWWlvlv\ channels, the primary vertex of the event is defined 
   as the reconstructed
  vertex with the highest $\sum\pt^2$, where $\pt$ is the magnitude  
  of the transverse momentum\footnote{ATLAS
  uses a right-handed coordinate system with its origin at the nominal
  interaction point (IP) in the centre of the detector, and the
  $z$-axis along the beam line.  The $x$-axis points from the IP to
  the centre of the LHC ring, and the $y$-axis points upwards.
  Cylindrical coordinates $(r,\phi)$ are used in the transverse plane,
  $\phi$ being the azimuthal angle around the beam line.  Observables
  labelled ``transverse'' are projected into the $x-y$ plane.
  The pseudorapidity is defined in terms of the polar angle $\theta$ as
  $\eta=-\ln\tan(\theta/2)$.} of each associated track; it is
  required to have at least three associated tracks with $\pt > 0.4\GeV$. 
  For the \hgg\ analysis a different primary vertex definition is used, 
   as described in Section~\ref{sec:hgg}.
 
  Muon candidates~\cite{Aad:2010yt} are formed by matching reconstructed 
  tracks in the inner detector (ID) with either complete tracks or 
  track segments reconstructed in the muon spectrometer (MS). 
   The muon acceptance is extended to the region $2.5<|\eta|<2.7$,
   which is outside the ID coverage, using tracks reconstructed in 
   the forward part of the MS.
  
  Electron candidates~\cite{Aad:2011mk} must have a
  well-reconstructed ID track pointing to a cluster of cells with energy depositions
  in the 
  electromagnetic calorimeter. The cluster should satisfy a set of identification
  criteria requiring the longitudinal and
  transverse shower profiles to be consistent with those expected for
  electromagnetic showers. Tracks associated with
  electromagnetic clusters are fitted using a Gaussian Sum
  Filter~\cite{GSFConf}, which allows 
  bremsstrahlung energy losses to be taken into account.
  The identification criteria described in Ref.~\cite{Aad:2011mk} have been modified 
  with time to maintain optimal performance as a function of pile-up, in particular for
  low-\pt\ electrons.
  
   The reconstruction, identification and trigger efficiencies for electrons
  and muons, as well as their energy and momentum scales and resolutions, 
  are determined using  
  large samples of $Z\to\ell\ell$, $W\to\ell\nu$ and $J/\psi\to\ell\ell$ 
   events~\cite{Aad:2011mk,ATLAS-CONF-2011-063}. The resulting uncertainties 
   are smaller than $\pm$1\% in most cases, one exception being the uncertainty
   on the electron selection efficiency 
   which varies between $\pm$2\% and $\pm$5\% as a function of \pt\ 
   and $\eta$. 
    
   Photon candidates~\cite{Aad:2010sp} are reconstructed and identified using shower shapes
   in the electromagnetic calorimeter, with or without associated conversion tracks, as described in 
   Section~\ref{sec:hgg}.  
  
  Jets~\cite{JES,JES2013} are built from topological 
  clusters~\cite{Lampl:1099735} 
  using the 
  anti-$k_t$ algorithm~\cite{Cacciari:2008gp} with
  a distance parameter $R=0.4$. They are typically
  required to have transverse 
  energies greater than 25~\GeV\ (30~\GeV) for $|\eta| < 2.4$ 
  ($2.4 \leq |\eta| < 4.5$), where the higher
 threshold in the forward region reduces the contribution from
 jet candidates produced by pile-up.
 To reduce this contribution further, jets within the ID acceptance 
 ($|\eta|<2.47$) are required to 
 have more than 25--75\% (depending on the pile-up conditions and Higgs boson decay mode)
  of the summed scalar \pt\ of their associated 
 tracks coming from tracks originating from the event primary vertex. Pile-up corrections based
 on the average event transverse energy density in the jet area~\cite{pileup} and
%~\cite{CSpile} 
 the number of reconstructed vertices in the data are also applied. 

  Jets originating from 
   $b$-quarks~\cite{btagnote2012l,btagnote2012b,BTagging}  
   are identified (``$b$-tagged") by combining 
  information from algorithms exploiting the impact parameter of tracks (defined as the distance 
  of closest approach to the primary vertex in the transverse plane), the presence of
  a displaced vertex,  and the reconstruction of $D$- and $B$-hadron decays.
 
  The missing transverse  momentum, \met~\cite{atlas_etmiss}, is the 
  magnitude of the negative vector sum of the \pt\ of muons, electrons, photons,
  jets and clusters of calorimeter cells  with $|\eta| < 4.9$ not associated 
  with these objects. 
   The uncertainty on the \met\ energy scale is obtained from the propagation of the uncertainties
   on the contributing components and thus depends on the considered final state. 	
   A track-based missing transverse momentum,
  \vMPT, is calculated as the negative vector sum of the transverse momenta 
  of tracks associated with the primary vertex. 
  
  The main sources of experimental uncertainty common to all the channels considered
  in this study are summarised in the top part of Table~\ref{tab:commonsys}.

%% file: Samples.tex
The SM Higgs boson production processes considered in these studies are 
gluon fusion ($gg\to H$, denoted ggF), vector-boson
fusion ($qq'\to qq'H$, denoted VBF), and Higgs-strahlung
($q\bar{q}'\to WH, ZH$, denoted $WH$/$ZH$ or jointly $VH$). 
 The small contribution from the associated production
with a \ttbar\ pair (\ttH, denoted $ttH$) is taken into
account in the \hgg\ and \hzz\ analyses. 
 \begin{table}[!htbp]
   \centering
   \caption{Event generators used to model the signal and the main background
     processes. ``PYTHIA'' indicates that PYTHIA6~\cite{pythia} and PYTHIA8~\cite{pythia8}
     are used for the simulations of 7\,TeV and 8\,TeV data, respectively.
     }
   \vspace{0.3cm}
   \scalebox{0.77}{
   \begin{tabular}{ll}
     \hline\hline
     Process & Generator \\
     \hline
     ggF, VBF & POWHEG~\cite{Alioli:2008tz,Nason:2009ai}+PYTHIA \\
     $WH$, $ZH$, $t\bar{t}H$ & PYTHIA  \\
     \hZZllll\  decay & PROPHECY4f~\cite{Bredenstein:2006rh,Bredenstein:2006ha} \\
     \hline
     $W$+jets, $Z/\gamma^{*}$+jets  & ALPGEN~\cite{alpgen}+HERWIG~\cite{Corcella:2000bw}, \\
                                    & POWHEG+PYTHIA, SHERPA~\cite{SHERPA} \\
     $\ttbar$, $tW$, $tb$           & MC@NLO~\cite{mcatnlo}+HERWIG \\
     $tqb$                          & AcerMC~\cite{Kersevan:2004yg}+PYTHIA6 \\
     $q\bar{q}\rightarrow WW$       & POWHEG+PYTHIA6  \\
     $gg\rightarrow WW$             & gg2WW~\cite{gg2WW, Kauer:2012hd}+HERWIG \\
     $q\bar{q}\to ZZ^*$               & POWHEG~\cite{Melia:2011tj}+PYTHIA \\
     $gg \to ZZ^*$                    & gg2ZZ~\cite{gg2ZZ, Kauer:2012hd}+HERWIG \\
     $WZ$                           & MadGraph~\cite{Alwall:2007st,Alwall:2011uj}+PYTHIA6, HERWIG \\
     $W\gamma$+jets                 & ALPGEN+HERWIG \\
     $W\gamma^{*}$                  & MadGraph~\cite{Gray:2011us}+PYTHIA6 for $m_{\gamma^*}< 7$~GeV\\
                                    & POWHEG+PYTHIA for $m_{\gamma^*} >7$~GeV \\
     $q\bar{q}/gg\to\gamma\gamma$   & SHERPA \\
     \hline\hline \\
   \end{tabular}
   } \label{tab:gen}
 \end{table}
 Samples of MC-simulated events 
are employed to model Higgs boson production and compute signal selection
efficiencies. The event generators are listed in Table~\ref{tab:gen}.
 Cross-section normalisations and other corrections ({\it e.g.}~Higgs 
 boson \pt\  spectrum)
%, some interference effects between signal and backgrounds) 
 are obtained from up-to-date calculations as described 
 in Refs.~\cite{paper2012ichep,YellowReport,LHCHiggsCrossSectionWorkingGroup:2012vm,HiggsXSWGWebPage,Georgi:1977gs,Djouadi:1991tka,Dawson:1990zj,
 Spira:1995rr,Harlander:2002wh,Anastasiou:2002yz,Ravindran:2003um,
 Aglietti:2004nj,Actis2008,Catani:2003zt,deFlorian:2012yg,Anastasiou:2012hx,Baglio:2010ae,
 deFlorian:2011xf,Bagnaschi:2011tu,Cahn:1983ip,Ciccolini:2007jr,Ciccolini:2007ec,Arnold:2008rz,Bolzoni:2010xr,
 Glashow:1978ab,Han:1991ia,Brein:2003wg,Ciccolini:2003jy,Kunszt:1984ri,Beenakker:2001rj,
 Beenakker:2002nc,Dawson:2002tg,Dawson:2003zu}.
Table~\ref{tab:cross} shows the production cross sections and the branching ratios for 
the final states considered in this study
for a Higgs boson with mass $m_H=125$\,GeV, while Table~\ref{tab:commonsys}
summarises the theoretical uncertainties on the expected signal common to all channels.  

Backgrounds are determined using data alone or a combination
of data and MC simulation, as discussed in Sections~\ref{sec:hgg}--\ref{sec:hww}. 
 The generators employed in most cases 
are also listed in Table~\ref{tab:gen}.
 To generate parton showers and their hadronisation, and to simulate the underlying 
event~\cite{MCTUNE2010,MCTUNEMC11,MCTUNEPY8},
PYTHIA6 (for 7\,TeV samples as well as for 8\,TeV samples produced
with MadGraph or AcerMC) or PYTHIA8 (for other 8\,TeV samples) are used.
Alternatively, HERWIG is employed, combined with the underlying
event simulation provided by JIMMY~\cite{jimmy}.
When PYTHIA6 or HERWIG are used,  
PHOTOS~\cite{Golonka:2005pn,Davidson:2010ew} is employed
to describe additional photon radiation from charged leptons.
 The small contributions from $Z^{(*)}$ and $W^{(*)}$ decays to electrons
 and muons through intermediate $\tau$-leptons are included in the signal
 and background generation. 

The following parton distribution function (PDF) sets are used in most cases: 
CT10 ~\cite{Lai:2010vv}
for the POWHEG, MC@NLO, gg2WW and gg2ZZ samples;
CTEQ6L1~\cite{cteq6} for the PYTHIA8, ALPGEN, AcerMC, MadGraph, HERWIG and SHERPA samples; and
MRSTMCal~\cite{mrst} for the PYTHIA6 samples.
In most cases, the generated MC samples are processed through a full
simulation~\cite{atlassim} of the ATLAS detector based on GEANT4~\cite{GEANT4}.
 Corrections obtained from measurements in the data are
applied to the simulation to account for small differences between data and simulation
in {\it e.g.}~the reconstruction of leptons, photons and jets.    
 The simulation also includes realistic modelling (tuned to the data) 
of the event pile-up from the same and nearby bunch crossings.

\begin{table}[!htbp]
  \centering
  \caption{SM Higgs boson cross sections (in pb)  
   at $\sqrt{s}$=8~(7)~TeV for $m_H=125$~GeV. The total values as well as the contributions
   from the individual production modes are listed. 
   The branching ratios to the final-state channels considered in this paper are also given
   (where $\ell$ stands for electron or muon), together with their relative uncertainty. 
    Up-to-date theoretical calculations are 
   used~\cite{YellowReport,LHCHiggsCrossSectionWorkingGroup:2012vm,HiggsXSWGWebPage,Djouadi:1997yw,
   Bredenstein:2006rh,Bredenstein:2006ha}.}
  \vspace{0.3cm}
  \scalebox{0.72}{
  \begin{tabular}{ll|ll}
    \hline\hline
%    \multicolumn{2}{l}{Cross section in pb at $\sqrt{s}$=8~(7)~TeV} \multicolumn{2}{c}{Branching ratios (relative uncertainty)}\\
     Cross section (pb)  &  & Branching ratio  & \\
     at $\sqrt{s}$=8~(7)~TeV &  & (relative uncertainty) & \\
    \hline
    ggF & 19.52 (15.32)   & \hWWlnln &  0.010   ($\pm$ 5\%)               \\
    VBF & 1.58 (1.22)     &  \hgg    &  2.28$\times10^{-3}$ ($\pm$ 5\%)   \\
    WH & 0.70 (0.57)      & \hzzllll &  1.25$\times10^{-4}$  ($\pm$ 5\%)  \\
    ZH & 0.39 (0.31)      &          &                        \\
    $t\bar{t}H$ & 0.13 (0.09)    &          & 		\\
    Total & 22.32 (17.51) &          &                         \\		   
    \hline\hline \\
  \end{tabular}
  } \label{tab:cross}
\end{table}

%% file: Hggcoupl2013.tex
 This channel is particularly sensitive to physics beyond the Standard Model 
 since the decay proceeds via loops (which 
 in the SM are dominated by $W$-boson exchange). 
 
% This channel is particularly sensitive to physics beyond the Standard Model 
% through possible anomalous contributions in particular
% to the \hgg\ decay loop (which 
% in the SM is dominated by $W$-boson exchange). 
   
 Events are required to have two high-\pt\ photons with invariant mass 
 in the range $100\,\text{--}\,160\gev$. The main background is continuum 
$\gamma\gamma$ production, with smaller
contributions from $\gamma+$jet and dijet processes.
Compared to the previously published results~\cite{paper2012ichep}, 
additional categories of 
events are introduced in the analysis of the 8~TeV data to 
increase the sensitivity to production through 
VBF or in association with a $W$ or $Z$ boson. 

\subsection{Event selection\label{hggsel}}

The data used in this channel are selected using a diphoton trigger~\cite{Aad:2012xs}
requiring two clusters formed from energy depositions in the electromagnetic 
calorimeter, with shapes compatible with electromagnetic showers. 
An $E_{\mathrm{T}}$ threshold of $20\GeV$ is applied to each cluster for the 7~TeV data,
 while at 8~TeV the thresholds are increased to $35\GeV$ on the leading 
(highest $E_{\mathrm{T}}$) and $25\GeV$ on the sub-leading (next-highest $E_{\mathrm{T}}$) cluster.
The trigger efficiency is larger than $99\%$ for events passing the final event selection.

%Events are required to contain at least one reconstructed vertex with at least two 
%associated tracks with $\pt>0.4$~GeV,
%as well as two photon candidates.

In the offline analysis, photon candidates are required to have $E_{\mathrm{T}} > 40 \gev$ 
and $30 \gev$ for the leading and sub-leading 
photon, respectively. Both photons must be reconstructed in the fiducial region 
$|\eta|<2.37$, excluding 
the calorimeter barrel/end-cap transition region $1.37\leq|\eta|<1.56$.

Photon candidates are required to pass tight identification criteria based mainly 
on shower shapes in the electromagnetic calorimeter~\cite{paper2012ichep}. They are 
classified as converted if they are associated with two tracks consistent with 
a $\gamma\to e^+e^-$ conversion process or a single track leaving no hit in the innermost 
layer of the inner detector, 
and as unconverted otherwise~\cite{ATL-PHYS-PUB-2011-007}.
%For the $7\TeV$ data, this information is combined using the same neural network technique as that of 
%the analysis described in Ref.~\cite{paper2012ichep}.
%For the $8\TeV$ data, cut-based criteria are used to ensure reliable photon performance for 
%recently-recorded data. 
 Identification efficiencies, averaged over $\eta$, range from $85\%$ to above $95\%$ for 
the $E_{\mathrm{T}}$ range under consideration.
Jets misidentified as photons are further rejected by applying calorimeter and track 
isolation requirements to 
the photon candidates. The calorimeter isolation is defined as the sum 
of the transverse energies of positive-energy topological clusters within a cone 
of size \DeltaR = 0.4 around the photon candidates,
excluding the core of the showers. It is required to be smaller 
than 4~GeV and 6~GeV for the 7~TeV and 8~TeV data, respectively. 
 The pile-up contribution is corrected on an event-by-event basis~\cite{Cacciari:2007fd}. 
 The track isolation, applied to the 8~TeV data only, is defined 
 as the scalar sum of the transverse momenta of all tracks with $\pt>1\,\gev$ associated
 with the diphoton production vertex (defined below) and lying within a cone 
 of size $\Delta R=0.2$ around the photon candidate; it
is required to be smaller than 2.6~GeV. Conversion tracks 
associated with either
photon candidate are excluded. 

For the precise measurement of the diphoton invariant mass (\mgg), as well as for the 
computation of 
track-based quantities ({\it e.g.}~track isolation, selection of jets associated with 
the hard interaction), the diphoton production vertex should be known precisely. 
 The determination of the vertex position along the beam axis
is based on so-called ``photon pointing", where the directions of the two photons, measured using 
the longitudinal and lateral  
segmentation of the electromagnetic calorimeter, are combined with a constraint from 
the average beam-spot position. For 
converted photons the position of the conversion vertex is also used. This 
technique alone is sufficient to ensure that the contribution of angular measurement uncertainties
to the diphoton invariant mass resolution is negligible. For a more precise identification of the
primary vertex, needed for the computation of track-based quantities, 
 this pointing information is combined with tracking information from each 
 reconstructed vertex: the $\Sigma p^2_{\rm T}$ for the 
 tracks associated with a given vertex and, for the 8~TeV data, the $\Sigma \pT$ of the 
 tracks and the azimuthal 
 angle between the transverse momentum of the diphoton system and that of 
the vector sum of the track \pTvec. A Neural Network (likelihood) discriminant is 
used for the 8~TeV (7~TeV) data.
The performance of this algorithm is studied using $Z \rightarrow ee$ decays, ignoring the tracks 
associated with the electrons and weighting the events so that the \pT\ and rapidity distributions 
of the $Z$ boson match those expected from the Higgs boson signal. The probability of 
finding a vertex within $0.3 \mm$ of the 
one computed from the electron tracks is larger than $75\%$.

 The photon energy calibration is obtained from a detailed 
simulation of the detector geometry and response, independently 
for converted and unconverted photons.
 The calibration is refined by applying $\eta$-dependent correction factors determined
 from studies of \Zee\ events in data~\cite{Aad:2011mk}: they 
range from $\pm 0.5\%$ to $\pm 1.5$\% depending on the pseudorapidity of the photon. 
 Samples of radiative $Z\to\ell\ell\gamma$ decays are used to verify the
photon energy scale. The energy response of the 
calorimeter shows a stability of better than $\pm0.1\%$ with time and various pile-up 
conditions.

  The signal efficiency of the above selections at 8~TeV 
 is estimated to be 37.5\% for a Higgs boson with $m_H=125$~GeV.
 
 The number of events in the diphoton mass region 100--160~GeV passing 
 this inclusive selection 
 is $23788$ in the 7~TeV data and $118893$ in the 8~TeV data. The fraction of  
 genuine $\gamma\gamma$ events, as estimated from data~\cite{Collaboration:2011ww},
 is $(75^{+3}_{-4})\%$.

\subsection{Event categorisation\label{hggcat}}

To increase the sensitivity to the overall Higgs boson signal, as well as to the specific 
VBF and VH production modes,
the selected events are separated into 14 mutually exclusive categories for further analysis, 
 following the order of preference listed below.

%The categories are presented below in order of decreasing priority, so that an event qualified for several categories is assigned to the one that was presented first. 
% here jet, lepton and MET object definitions ?

{\em Lepton category} (8~TeV data only): This category targets mainly $VH$ 
events where the $W$ or $Z$ bosons decay to charged leptons. An isolated electron 
 (${\ensuremath{E_{\mathrm{T}}}}>$15~GeV)  or muon 
(\pt\ $>$ 10~GeV) candidate is required. To remove 
contamination from $Z\gamma$ production with $Z\rightarrow ee$, electrons forming 
an invariant mass with either photon in the range $84 \gev\ < m_{e\gamma} < 94 \gev$  
are not considered.

{\em $E_T^{miss}$  category} (8~TeV data only): This category targets mainly $VH$ events 
with $W \rightarrow \ell\nu$ or $Z \rightarrow \nu\nu$.
 An \MET\ significance (defined as $\MET/\sigma_{\MET}$, 
 where in this case $\sigma_{\MET}=0.67\gev^{1/2}\sqrt{\Sigma E_T}$ with
$\Sigma E_T$ being the event total transverse energy)
greater than five is required, corresponding to $\MET > ~70\,\text{--}\,100 \gev$ depending on $\Sigma E_T$.
%Events where either photon also passes electron identification requirements are not selected. 

{\em Low-mass two-jet category} (8~TeV data only): To select $VH$ events where the $W$ or $Z$ boson 
decays hadronically, a pair of jets with invariant mass in the range
$60\gev\ < m_{jj} < 110 \gev$ is required. 
To reduce the ggF contamination, the pseudorapidity  difference between the 
dijet and diphoton systems is required to be $|\Delta\eta_{\gamma\gamma,jj}| < 1$, and 
 the component of the diphoton transverse momentum 
orthogonal to the diphoton thrust axis in the transverse 
plane\footnote{$p_{\rm{Tt}} = |(\vec{p}_\mathrm{T}^{\gamma_1} + 
\vec{p}_\mathrm{T}^{\gamma_2}) \times \hat{\bf t}|$, 
where $\hat{\bf t} = \frac{ {\vec{p}_\mathrm{T}^{\gamma_1}} -{\vec{p}_\mathrm{T}^{\gamma_2}} }
{ |{\vec{p}_\mathrm{T}^{\gamma_1}} 
- {\vec{p}_\mathrm{T}^{\gamma_2}}|}$ denotes the thrust axis in the transverse plane, and
${\vec{p}_\mathrm{T}^{\gamma_1}}$, ${\vec{p}_\mathrm{T}^{\gamma_2}}$ are the transverse 
momenta of the two photons.}~\cite{PTT_OPAL,PTT_ZBoson} is required to satisfy $\ptt > 70 \gev$. 

\begin{figure}[!ht]
\centering
\includegraphics[width=.5\textwidth]{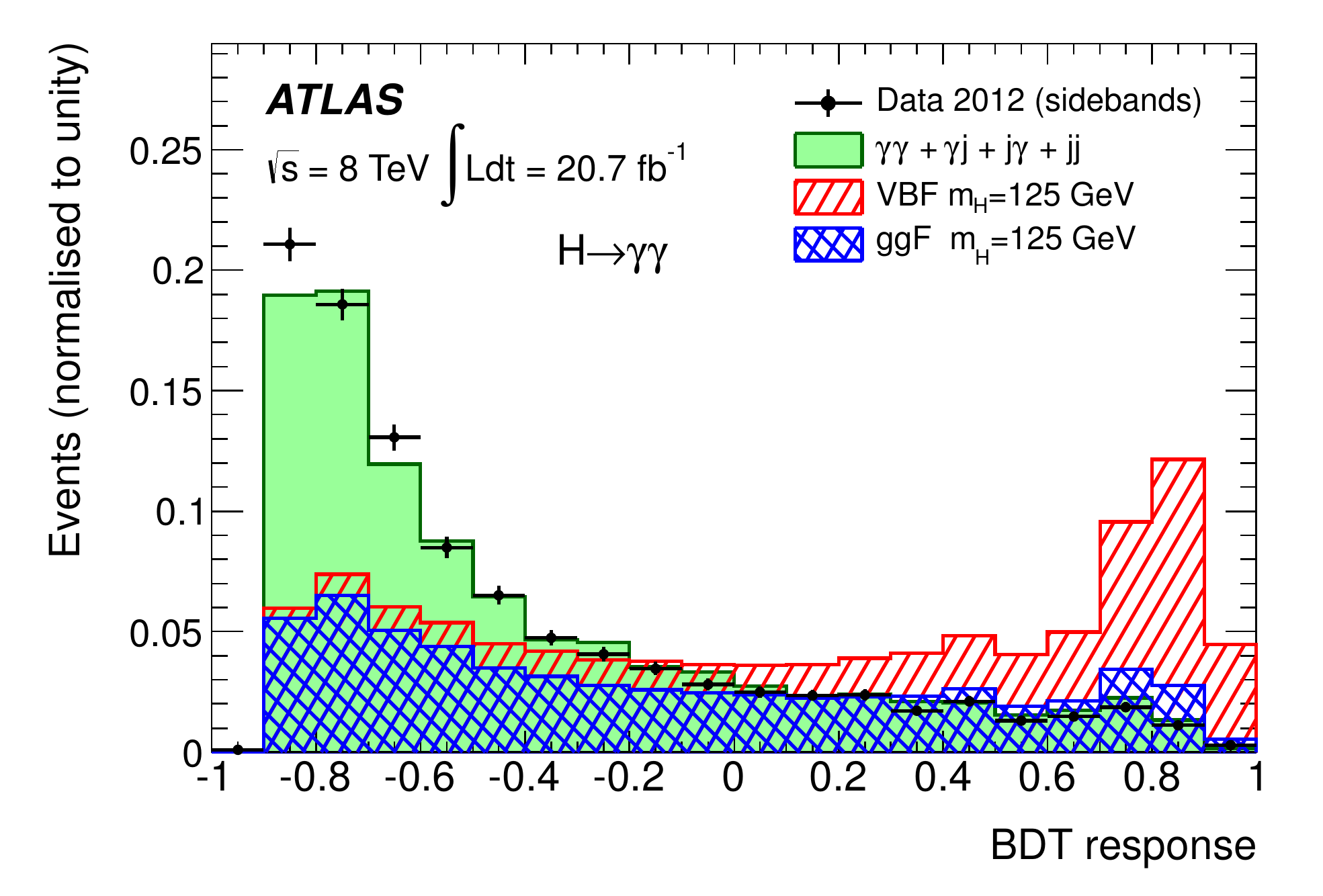}
\caption{Distribution of the VBF BDT response after applying the selection
of the inclusive analysis and requiring in addition the presence of two jets with
$|\Delta\eta_{jj}| > 2$ and $|\eta^*|<5$.
The data in the signal sidebands ({\it i.e.}~excluding the \mgg\ region 120--130~GeV), the
expected background, and the expected signal from VBF and ggF production are shown. They are all 
normalised to unity except ggF, 
which is normalised to the ratio between the numbers of ggF and VBF events passing 
the selection described above.} 
\label{fig:hgg:BDT} 
\end{figure}

{\em High-mass two-jet categories}: These categories are designed to select events produced 
through the VBF process, which is
characterised by the presence of two forward jets with little hadronic activity in the central 
part of the detector. Jets are reconstructed as described in Section~\ref{sec:physobj}. 
 The selection for the 8 TeV 
 data is based on a multivariate technique using a Boosted Decision Tree (BDT), whose input quantities are: 
the pseudorapidities of the two jets ($\eta_{j1}$, $\eta_{j2}$) and their separation in $\eta$;
 the invariant mass of the dijet system; the difference 
 $\eta^* = \eta_{\gamma\gamma} - (\eta_{j1} + \eta_{j2})/2$, 
 where $\eta_{\gamma\gamma}$ is the pseudorapidity of the diphoton system; 
 the minimal radial distance (\DeltaR) of any jet--photon pair; and the difference \dphiggjj\ between 
 the azimuthal angles of the diphoton and dijet momenta. %explain the relation to jet veto ?
The BDT training is performed using a signal sample, as well as a background sample composed of 
simulated $\gamma\gamma$ events combined with $\gamma j$ and $jj$ components obtained from data. 
%The relative fraction of each component is fixed to the fractions measured in 
%data reported in Section~\ref{bkg} below. 
The BDT response distributions for data and simulation are 
shown in Fig.~\ref{fig:hgg:BDT}.
The BDT output is used to define two high-mass two-jet categories: a ``tight" category corresponding 
to ${\rm BDT } \ge 0.74$, and a ``loose" category for $0.44 \le {\rm BDT }< 0.74$. 
For the 7~TeV data, the same cut-based selection as described in 
Ref.~\cite{paper2012ichep} is applied, namely $m_{jj} > 400 \gev$, 
$|\Delta\eta_{jj}| > 2.8$ and $|\dphiggjj| > 2.8$.

{\em Untagged categories}: Events not selected in any of the above categories 
(corresponding to more than 90\% of the expected signal, dominated by ggF production) are 
 classified in nine 
additional categories according to the 
properties of their diphoton system. Events with both photons unconverted 
are classified into 
{\em unconverted central} if $|\eta|<0.75$ for both photons, and 
 {\em unconverted rest} otherwise. Events with at 
least one converted photon are similarly separated into {\em converted central} if $|\eta|<0.75$ for both photons, {\em converted transition} if 
$1.3 < |\eta| < 1.75$ for either photon, and {\em converted rest} otherwise. Finally, all untagged 
categories except {\em converted transition} are 
split into {\em low \ptt} and {\em high \ptt} sub-categories by a cut at $\ptt = 60 \gev$.
 This classification  is motivated by differences in mass resolution and signal-to-background 
 ratio for the various categories.  
 
 The use of the 14 categories improves the sensitivity of the analysis by about 40\% compared to the 
 inclusive analysis.

\subsection{Background estimation\label{hggfit}}

The background is obtained from fits to the diphoton mass spectrum in the data
over the range 100--160~GeV 
after the full selection.  The procedure, the choice of the analytical forms 
for the background and the 
determination of the corresponding uncertainties follow the method described 
in Ref.~\cite{paper2012ichep}. Depending on the category, the analytical form
 is either a fourth-order Bernstein polynomial~\cite{Bernstein}
(used also for the inclusive sample), an exponential of a second-order polynomial, or 
a single exponential. In these fits, the Higgs boson signal is described 
by the sum of a Crystal Ball function~\cite{oreglia} 
for the core of the distribution and 
a Gaussian function for the tails.

\subsection{Systematic uncertainties\label{sec:hggsyst}}

Systematic uncertainties can affect the signal yield, 
the signal fractions in the various categories (with possible migrations between them), the 
signal mass resolution and the mass measurement. The main 
sources specific to the \hgg\ channel are listed in Table~\ref{tab:hgg:systs}, while sources
in common with other decay channels are summarised in Section~\ref{sec:physobj}  
and Table~\ref{tab:commonsys}. The uncertainties described below are those 
affecting the 8~TeV analysis (see Ref.~\cite{paper2012ichep} for the 7~TeV analysis).
\begin{table}[htbp]
  \begin{center}
   \caption{For $m_H=125$~GeV and the 8~\TeV\ data analysis, the impact of the main sources 
    of systematic uncertainty specific to the \hgg\ channel 
    on the signal yield, 
    event migration between categories and mass measurement and resolution. 
     Uncertainties common to all channels are listed in
    Table~\ref{tab:commonsys}.     
     The $\pm$ and $\mp$ signs indicate anticorrelations between categories.}
\vspace{0.1cm}
\scalebox{0.70}{
    \begin{tabular}{lc}
      \hline \hline
       Source            &  Uncertainty (\%) \\
      \hline\hline
         \multicolumn{2}{r}  {on signal yield  }  \\   
\hline	    
%      \multicolumn{2}{l}{Uncertainty (\%) on signal yield } \\
      \quad Photon identification   & $\pm 2.4$   \\
      \quad Trigger                 & $\pm 0.5$   \\      
      \quad Isolation               & $\pm 1.0$   \\
      \quad Photon energy scale     & $\pm 0.25$  \\
      \quad ggF (theory), tight high-mass two-jet cat. & $\pm 48$  \\ 
      \quad ggF (theory), loose high-mass two-jet cat. & $\pm 28$  \\ 
      \quad ggF (theory), low-mass two-jet cat.      & $\pm 30$ \\
      \quad Impact of background modelling         & $\pm (2\,\text{--}\,14)$, cat.-dependent \\
      \hline
%      \multicolumn{2}{l}{Uncertainty on poulation of categories (migration, \%)} \\
        \multicolumn{2}{r} { on category  population (migration) } \\   
\hline	    	 
      \quad Material modelling & $- 4$ (unconv), $+ 3.5$ (conv)\\
      \quad \pT\ modelling & $\pm 1$ (low-\ptt), \\
                     & $\mp (9\,\text{--}\,12)$ (high-\ptt, jets), \\
                     & $\pm (2\,\text{--}\,4)$ (lepton, \MET) \\
      \quad \dphiggjj, $\eta^*$ modelling in ggF       & $\pm (9\,\text{--}\,12)$, $\pm (6\,\text{--}\,8)$ \\
      \quad Jet energy scale and resolution & $\pm (7\,\text{--}\,12)$ (jets), \\
          & $\mp (0\,\text{--}\,1)$ (others) \\
      \quad Underlying event two-jet cat. & $\pm 4$ (high-mass tight),  \\
          &  $\pm 8$ (high-mass loose), \\
          &  $\pm 12$ (low-mass) \\
%      \MET & $\pm 66$ (ggF), $\pm 31$ (VBF), $\pm 1$ (others) \\
      \quad \MET & $\pm 4$ (\MET\ category) \\
      \hline
%      \multicolumn{2}{l}{ Uncertainty on mass measurement and resolution (\%)} \\
          \multicolumn{2}{r} {  on mass scale and resolution } \\       	
\hline	   
      \quad Mass measurement      & $\pm 0.6$, cat.-dependent \\ 
      \quad Signal mass resolution         & $\pm (14\,\text{--}\,23)$, cat.-dependent \\
      \hline \hline
    \end{tabular}
    }
\label{tab:hgg:systs}
\end{center}
\end{table}

{\em Signal yield}: Relevant experimental uncertainties on the signal yield come from the
knowledge of the luminosity (Table~\ref{tab:commonsys}) and the
photon identification efficiency. The latter is estimated by comparing
the efficiencies obtained using MC simulations and several data-driven methods: 
 $Z \rightarrow ee$ events with a simulation-based extrapolation from 
electrons to photons, an isolation sideband technique using an inclusive photon sample, 
 and photons from $Z \rightarrow \ell\ell\gamma$ radiative decays. Owing to several
 analysis improvements and the large size of the
 8~TeV data sample, the resulting uncertainty is significantly 
 reduced compared 
 to that reported in Ref.~\cite{paper2012ichep}
 and amounts to $\pm2.4\%$. Smaller experimental 
 uncertainties come from the knowledge of the trigger efficiency, 
  the impact of the photon isolation requirement and the photon energy scale.
 In addition to the theoretical uncertainties on inclusive Higgs boson production listed in 
 Table~\ref{tab:commonsys}, the ggF contribution to the two-jet 
 categories is subject to large uncertainties (Table~\ref{tab:hgg:systs}) due to missing higher-order 
 corrections; they are estimated using the method described in Ref.~\cite{TackmannGangal} and 
  the MCFM~\cite{MCFM2010} generator calculations. 
% For the high-mass two-jet categories the uncertainty on the BDT-based selection is 
% applied on an event-by-event basis.
 Finally, the background modelling contributes an uncertainty between $\pm$2\% and $\pm$14\% 
 depending on the category.

{\em Event migration}: 
Mis-modelling of the detector material could cause event migration between the unconverted and converted
photon categories in the simulation. The uncertainty is obtained from MC samples produced with variations of the
material description. The uncertainty in the population of the 
\ptt\ categories due to the description of the Higgs boson \pT\ spectrum 
is determined by varying the QCD scales 
and PDFs used in the $HqT$ program~\cite{deFlorian:2011xf}.
Uncertainties on the modelling of two-jet variables for the ggF process, in particular
\dphiggjj\ and $\eta^*$, affect the contribution of ggF events to the high-mass two-jet categories.
They are estimated by comparing the baseline POWHEG generator
with SHERPA and MCFM. Uncertainties on the jet energy scale and 
resolution affect the selection of jets used in some 
category definitions, thereby causing migration between jet-based and other categories.
 The uncertainty due to the modelling of the underlying event is estimated by comparing 
simulations with and without multi-parton interactions.
 Uncertainties on the \MET\ reconstruction are assessed by varying the transverse energies
 of its components (photons, electrons, jets, soft energy deposits) within their
 respective uncertainties.

{\em Mass measurement and mass resolution}: The measurement of the Higgs boson mass 
in the \hgg\ channel
is discussed in Section~\ref{mass-mu}. Uncertainties on the diphoton mass scale 
come from the following sources: the calibration of the electron energy scale (obtained 
from $Z \rightarrow ee$ events); the uncertainty on its extrapolation to the energy scale of photons, 
dominated by the description of the detector material; and the 
knowledge of the energy scale of the presampler detector located in front of the
electromagnetic calorimeter. The total uncertainty amounts to $\pm$0.55\% 
 (corresponding to $\pm$0.7~GeV). 
 The mass resolution, obtained from the Crystal Ball function used in the fits described
 in Section~\ref{hggfit}, ranges from $1.4\gev$ to $2.5\gev$ depending on the category.
 The main uncertainties come from the calorimeter energy scale and the
extrapolation from the electron to the photon response. 
Smaller contributions arise from pile-up and the primary vertex selection.

\subsection{Results}
\begin{figure}
\begin{center}
\hspace*{-0.5cm}\includegraphics[width=.52\textwidth]{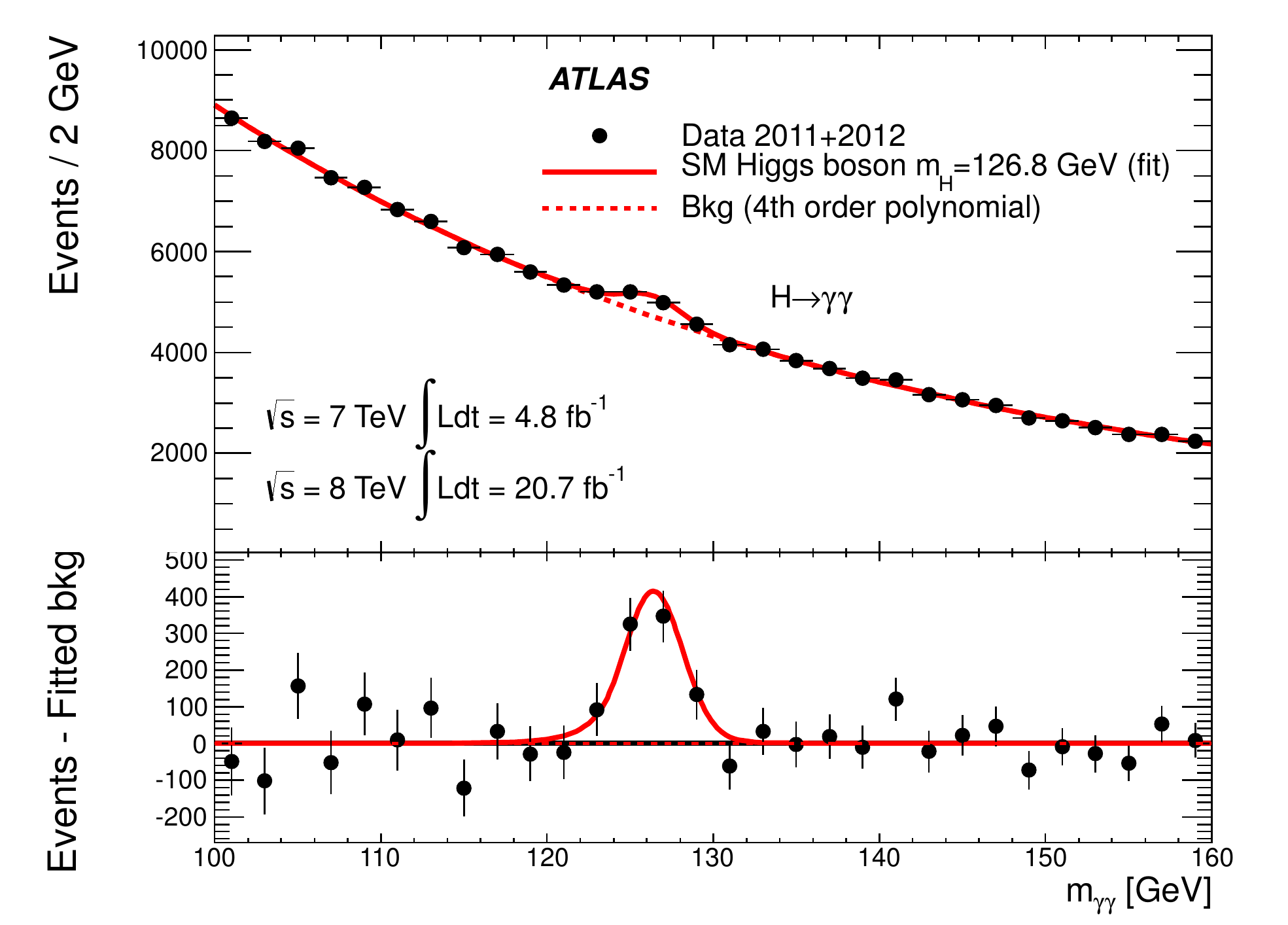}
\end{center}
\caption{Invariant mass distribution of diphoton candidates after all selections of the inclusive analysis
  for the combined 7~TeV and 8~TeV data.
  The result of a fit to the data with the sum of a SM Higgs boson signal (with $m_H=126.8$~\GeV 
  and free signal strength) and background is superimposed.
  The residuals of the data with respect to the fitted background are displayed in the lower panel.
}
\label{fig:hgg:mgg}
\end{figure}

The diphoton invariant mass distribution after
selections for the full data sample is shown in Fig.~\ref{fig:hgg:mgg}. 
 The data are fit by categories, using background shapes (see Section~\ref{hggfit}), as well as parameters
 for the Crystal Ball and Gaussian functions describing the signal, specific to each category.  
 At the maximum deviation from the
background-only expectation, which occurs for $m_H\sim126.5\GeV$, the significance of the observed peak
is 7.4$\sigma$ for the combined 7~\TeV\ and 8~\TeV\ data (compared with 4.3$\sigma$ expected from 
SM Higgs boson production at this mass), 
 which establishes a discovery-level signal in the $\gamma\gamma$ channel alone. 
%A summary of the categories and their properties is shown in Fig.~\ref{cats}. FG: cannot refer to aux material
\begin{table}[htbp]
\begin{center}
\caption{For the \hgg\ analysis of the $\sqrt{s}=8\TeV$ data, the numbers of events observed in the 
data~($N_D$), the numbers of background events~($N_B$) estimated from fits to the data, 
and the expected SM Higgs boson signal~($N_S$) for 
$m_H=126.8$~GeV, split by category. All numbers are given in a mass window 
centred at $m_H=126.8$~GeV and containing 90$\%$ of the expected signal (the size of this 
window changes from category to category and for the inclusive sample). The predicted
numbers of signal events in each of the ggF, VBF, $WH$, $ZH$ and $t\bar{t}H$ processes are also given.}
% The statistical uncertainty on $N_S$ is less than 1\%.}
% \end{center}
\label{tab:SignalNumbers}
\vspace{0.1cm}
\scalebox{0.6} {
\hspace*{-0.3cm}\begin{tabular}{l|r|r|rrrrrr}
\hline\hline
Category                   &  $N_D$   & $N_B$ & $N_S$ & ggF & VBF & $WH$  & $ZH$ & $t\bar{t}H$\\
\hline
Untagged                   &  14248   & 13582 & 350 & 320   &   19    & 7.0   &     4.2 &   1.0 \\
Loose high-mass two-jet    &     41   & 28    & 5.0 & 2.3   &  2.7    & $<0.1$ & $<0.1$ &  $<0.1$ \\
Tight high-mass two-jet    &     23   & 13    & 7.7 & 1.8   &  5.9    & $<0.1$ & $<0.1$ &  $<0.1$ \\
Low-mass two-jet           &     19   & 21    & 3.1 & 1.5   &   $<0.1$ & 0.92  &    0.54 &   $<0.1$ \\
\MET\ significance         &     8    & 4     & 1.2 & $<0.1$ &  $<0.1$ & 0.43  &    0.57 &  0.14 \\
Lepton                     &     20   & 12    & 2.7 & $<0.1$ &  $<0.1$  & 1.7   &    0.41 &  0.50 \\
\hline
All categories~(inclusive) & 13931   & 13205 & 370 & 330 &  27 &   10 &   5.8 &   1.7 \\
\hline\hline
\end{tabular}
}
\end{center}
\end{table}
Table~\ref{tab:SignalNumbers} lists the observed number of events in the main
categories, the estimated background from fits to 
the data (described in Section~\ref{hggfit}), and
the predicted signal contributions from the various production processes. 

Additional interpretation of these results is presented in Section~\ref{sec:results}.

%% file: H4lcoupl2013.tex
\renewcommand{\labelitemi}{$-$}
\newcommand\htollllp{$H\to ZZ^{*}\to 4\ell$}
\newcommand\htollllbrief{$H\to ZZ^{*}\to 4\ell$}
\newcommand\progname[1]{{\sc #1}}
\newcommand\pval{\ensuremath{p_0}}
\def\brocket#1{\left\langle #1 \right\rangle}
\def\Figref#1{Figure~\ref{#1}} % At the start of a sentence.
\def\figref#1{Fig.~\ref{#1}}   % Within a sentence.
\def\secref#1{Section~\ref{#1}}
\newcommand{\tabscript}[3]{%
  \setlength{\fboxrule}{0pt}%
  \fbox{\ensuremath{#1^{#2}_{#3}}}%
}
\newcommand\candtwelvemu{22~}
\newcommand\candtwelveemu{19~}
\newcommand\candtwelvemue{14~}
\newcommand\candtwelvee{14~}
\newcommand\candtwelvemix{33~}
\newcommand\candtwelvetotal{79~}
\newcommand\bkgtwelveexp{$XX\pm X$~}
\newcommand\candmu{33~}
\newcommand\candmix{32~}
\newcommand\cande{21~}
\newcommand\candtotal{86~}
\newcommand\candmuexp{22.4~$\pm$~X.X~}
\newcommand\candmixexp{33.7~$\pm$~X.X~}
\newcommand\candeexp{15.3~$\pm$~X.X~}
\newcommand\bkgexp{71.2~$\pm$~X.X~}
\newcommand\lumiFourMuona{\ensuremath{4.8~\ifb}}
\newcommand\lumiFourElectrona{\ensuremath{4.9~\ifb}}
\newcommand\lumiTwoMuonTwoElectrona{\ensuremath{4.8~\ifb}}
\newcommand\lumiFourMuonb{\ensuremath{5.8~\ifb}}
\newcommand\lumiFourElectronb{\ensuremath{5.9~\ifb}}
\newcommand\lumiTwoMuonTwoElectronb{\ensuremath{5.8~\ifb}}
\newcommand\sensitivityExpected{0.XX}
\newcommand\sensitivityObserved{0.XX}
\newcommand\sensitivityMass{\ensuremath{200\:\gev}}

\newcommand\pvaluelowexp{10.6\%}
\newcommand\sigmalowexp{\ensuremath{1.3}}
\newcommand\pvaluelow{1.6\%}
\newcommand\sigmalow{\ensuremath{2.1}}
\newcommand\pvaluelowm{\ensuremath{125\gev}}

\newcommand\pvaluelowmOld{\ensuremath{242\gev}}
\newcommand\pvaluelowOld{0.5\%}
\newcommand\sigmalowOld{2.6}

\newcommand\pvaluehighmOld{\ensuremath{125\gev}}
\newcommand\pvaluehighOld{1.1\%}
\newcommand\sigmahighOld{2.3}

\newcommand\pvaluelowmNew{\ensuremath{125.5\gev}}
\newcommand\pvaluelowNew{0.4\%}
\newcommand\sigmalowNew{2.7}

\newcommand\pvaluehighmNew{\ensuremath{266\gev}}
\newcommand\pvaluehighNew{3.5\%}
\newcommand\sigmahighNew{1.8}

\newcommand\pvaluelowmComb{\ensuremath{125\gev}}
\newcommand\pvaluelowComb{0.018\%}
\newcommand\sigmalowComb{3.6}

\newcommand\pvaluehighmComb{\ensuremath{266\gev}}
\newcommand\pvaluehighComb{1.9\%}
\newcommand\sigmahighComb{2.1}

\newcommand\globalSigmaLowComb{2.5}
\newcommand\globalpvalue{0.65\%}
%%%%%%%%%%%%%%%%%%%%%%%%%%%%%%

\newcommand\pvaluehighexp{0.14\%}
\newcommand\sigmahighexp{\ensuremath{3.0}}

\newcommand\pvaluehigh{1.3\%}
\newcommand\sigmahigh{\ensuremath{2.2}}
\newcommand\pvaluehighm{\ensuremath{244\gev}}
\newcommand\pvaluehightwoexp{7.1\%}

\newcommand\sigmahightwoexp{\ensuremath{1.5}}
\newcommand\pvaluehightwo{1.8\%}
\newcommand\sigmahightwo{\ensuremath{2.1}}
\newcommand\pvaluehighmtwo{\ensuremath{500\gev}}

\newcommand\excludedrangeaBrief{131--162}
\newcommand\excludedrangebBrief{170--460}
\newcommand\excludedrangeexpaBrief{124--164}
\newcommand\excludedrangeexpbBrief{176--500}
\newcommand\excludedranges{$\excludedrangeaBrief\,\gev$ and $\excludedrangebBrief\,\gev$}
\newcommand\excludedrangesexp{$\excludedrangeexpaBrief\,\gev$ and $\excludedrangeexpbBrief\,\gev$}

\newcommand\excludedrangesJan{$134\!-\!156\,\gev$, $182\!-\!233\,\gev$, $256\!-\!265\,\gev$ and $268\!-\!415\,\gev$}
\newcommand\excludedrangesexpJan{$136\!-\!157\,\gev$ and $184\!-\!400\,\gev$}
\newcommand\lumiAverageJan{\ensuremath{4.8~\ifb}}

\newcommand\lumia{\ensuremath{4.6~\ifb}}
\newcommand\lumib{\ensuremath{20.7~\ifb}}
\newcommand\lumiaPrev{\ensuremath{4.9~\ifb}}
\newcommand\lumibPrev{\ensuremath{13.0~\ifb}}
\newcommand\zzstar{$ZZ^{*}$}

    Despite the small branching ratio, this channel
   provides good sensitivity to Higgs boson studies, {\it e.g.}~to the coupling 
   to $Z$ bosons, mainly because of the large signal-to-background ratio. 
   
  Events are required to have two pairs of same-flavour, opposite-charge,
  isolated leptons: $4e$, $2e2\mu$, $2\mu2e$, $4\mu$ (where final states with
  two electrons and two muons are ordered by
  the flavour of the dilepton pair with mass closest to the $Z$-boson mass).    
  The largest background comes from continuum
  $(Z^{(*)}/\gamma^{*}) (Z^{(*)}/\gamma^{*})$ production, referred to
  hereafter as $ZZ^{*}$.  Important 
  contributions arise also from $Z+\rm{jets}$ and $t\bar{t}$
  production, where two of the charged lepton candidates can come from decays
  of hadrons with $b$- or $c$-quark content, misidentification
  of light-quark jets, and photon conversions.

The analysis presented here is largely the 
same as that described in Ref.~\cite{new_ZZ_council2012} 
with only minor changes.  
 The electron identification is tightened in the 8\,\tev\ data to
improve the background rejection for final states with a pair of
electrons forming the lower-mass $Z^{*}$ boson. 
The mass measurement uses a constrained fit
to the \Zboson\ mass to improve the resolution. The lepton pairing is
modified to reduce the mis-pairing in the 4$\mu$ and $4e$ final
states, and the minimum requirement on the mass of the second
$Z^{*}$ boson is relaxed.  Final-state radiation (FSR) is included 
in the reconstruction of the first $Z^{(*)}$ in events containing muons. 
Finally, a classification which separates Higgs boson candidate events into
ggF--like, VBF--like and VH--like categories is introduced. 
  
\subsection{Event selection}

The data are selected using single-lepton or dilepton triggers. 
The \pt~threshold of the single-muon trigger is 
24~GeV (18~GeV) in 2012 (2011) and 
the \et~threshold of the single-electron trigger is 
24~GeV (20--22~GeV).
The dielectron trigger threshold is $\et=12$\,\gev\ 
and the dimuon trigger threshold is $\pt=13$\,\gev\ (10\,\gev\ in 2011) 
for both leptons. In addition, an asymmetric dimuon trigger and
electron--muon triggers are used as described in Ref.~\cite{new_ZZ_council2012}.
 The efficiency for events passing the offline analysis cuts to be
selected by at least one of the above triggers is between 97\% and 100\%.

 Muon and electron candidates are reconstructed as described in Section~\ref{sec:physobj}.
 In the region $|\eta|<0.1$, which has limited MS coverage, 
 ID tracks with $\pt>15$\,\gev\ are identified as muons 
 if their calorimetric energy deposits are consistent with a minimum ionising particle.  
 Only one muon per event is allowed to be reconstructed either 
 in the MS alone or without MS information.
 For the 2012 data, the electron requirements are tightened in the transition region between 
the barrel and end-cap calorimeters ($1.37<|\eta|<1.52$), and the pixel-hit
requirements are stricter to improve the rejection of photon conversions.

Each electron (muon) must satisfy $\et>7$\,\gev\ ($\pt>6$\,\gev) and
be measured in the pseudorapidity range $\left|\eta \right|<2.47$
($\left|\eta\right|<2.7$). The highest-\pt~lepton in the quadruplet
must satisfy $\pt >20$\,\gev, and the second (third) lepton 
must satisfy $\pt >15$\,\gev\ ($\pt >10$\,\gev). To reject cosmic rays, muon tracks 
are required to have a transverse impact parameter of less than $1\,{\rm mm}$.

Multiple quadruplets within a single event are possible. For each quadruplet, 
the same-flavour, opposite-charge lepton pair
with invariant mass closest to the $Z$-boson mass ($m_{\Zboson}$)
is referred to as the leading lepton pair. Its invariant mass,
denoted by $m_{12}$, is required to be between $50\,\gev$~and
$106\,\gev$. The invariant mass of the other (sub-leading) lepton pair, $m_{34}$, is
required to be in the range $m_{\rm{min}}<m_{34}<115\,\gev$. 
The value of $m_{\rm{min}}$ is 12\,\gev\ for a reconstructed four-lepton mass
$m_{4\ell}<$ 140~GeV, rises
linearly to 50\,\gev\ at $m_{4\ell}=$ 190\,\gev, and remains constant
for higher masses. 
If two or more quadruplets satisfy the above requirements, the one with  $m_{34}$
closest to the \Zboson-boson mass is selected.  For further analysis, events
are classified in four sub-channels, $4e$, $2e2\mu$, $2\mu2e$, $4\mu$.

The \Zjets\ and $t\bar{t}$~background contributions are
reduced by applying requirements on the lepton transverse impact parameter divided by its
uncertainty, $|d_0|/\sigma_{d_0}$. This ratio must be smaller than 3.5 
for muons and smaller than 6.5 for electrons (the electron impact parameter 
is affected by bremsstrahlung and thus its distribution has longer tails). 
 In addition, leptons must satisfy isolation requirements based on tracking and calorimetric 
information, similar to those described in Section~\ref{hggsel}, 
as discussed in Ref.~\cite{paper2012ichep}.

The impact of FSR photon emission on the reconstructed invariant mass 
is modelled using the MC simulation (PHOTOS), which reproduces
the rate of collinear photons with \et~$>1.3$\,\gev\ in $Z\to\mu\mu$ decays in data to 
$\pm$5\%~\cite{ATLAS-CONF-2012-143}. Leading muon pairs 
with 66\,\gev~$<m_{12}<$~89\,\gev\ are corrected 
 for FSR by including any reconstructed photon 
 with \et\ above 1\,\gev\ lying close (typically within $\Delta R<$ 0.15) to the muon tracks,
 provided that the corrected $m_{12}$ satisfies $m_{12} <$~100\,\gev. The MC simulation
 predicts that about 4\% of all $H\rightarrow ZZ^{*}\rightarrow 4\mu$ candidate events 
 should have this correction.

 For the $8\,\tev$ data,
 the signal reconstruction and selection efficiency for a SM Higgs boson with $\mH=125\,\gev$
 is $39$\% for the $4\mu$ sub-channel,
$26$\% for the $2e2\mu$/$2\mu2e$ sub-channels and 
$19$\% for the $4e$ sub-channel. 

 The final discriminating variable in this
analysis is the $4\ell$ invariant mass. Its resolution, which is improved 
by typically 15\% by applying a $Z$-mass constrained 
kinematic fit to the leading lepton pair,  is about $1.6\,\gev$, 
$1.9\,\gev$ and
$2.4\,\gev$ for 
the $4\mu$, $2e2\mu$/$2\mu2e$ and $4e$ sub-channels, respectively, and for $m_{H}=125\,\gev$.

\subsection{Event categorisation}
\label{sec:eventCategorisation2012}

To enhance the sensitivity to the individual production modes, 
events passing the above selection are assigned to one of three 
categories, named VBF--like, VH--like, and ggF--like.
Events are VBF--like if the two highest \pt~jets are separated by more than 
three units in pseudorapidity and have an invariant mass greater than 350\,\gev.  
Events that do not qualify as VBF--like are considered for the VH--like category. 
They are accepted in this category if 
they contain an additional lepton ($e$ or $\mu$)
 with $\pt >$ 8\,\gev, satisfying the same requirements as the four leading leptons. 
 The remaining events are assigned to the ggF--like category. 
 No classification based on the $4\ell$ flavour 
 is made in the VBF--like and VH--like categories.
Higgs boson production through VBF and VH is expected to account for
about 60\% and 70\% of the total signal events in the VBF--like and
VH--like categories, respectively. The signal-to-background ratio in
the signal peak region is about five for the VBF--like category, about three
for the VH--like category, and about 1.5 for the inclusive analysis.

\subsection{Background estimation}

The expected background yield and composition are estimated using the MC simulation 
for $ZZ^{*}$ production, and  methods based on control regions (CRs) from 
data for the $Z+\rm{jets}$ and $t\bar{t}$ 
processes~\cite{paper2012ichep}. The transfer factors used to extrapolate the background 
yields from the CRs to the signal region are obtained from the MC simulation
and cross-checked with data.
Since the background composition depends on the flavour of the sub-leading lepton pair, 
different approaches are followed for the $\ell\ell+\mu\mu$ and the $\ell\ell+ee$ final states.

The reducible $\ell\ell+\mu\mu$ background is dominated by $t\bar{t}$ and $Z+{\rm jets}$ 
(mostly $Zb\bar{b}$) events. A CR is defined by removing the isolation 
requirement for the muons of the sub-leading pair, and by requiring that at least 
one of them fails the transverse impact parameter selection. 
This procedure allows the $t\bar{t}$ and $Z+{\rm jets}$ backgrounds to be estimated 
simultaneously from a fit to the $m_{12}$ distribution.

To determine the reducible $\ell\ell+ee$ background, a CR is formed by 
relaxing the selection criteria for the electrons of the sub-leading pair: 
each of these electrons is then classified as ``electron--like" or ``fake--like" 
based on requirements 
on appropriate discriminating variables~\cite{Aad:2011rr}. The numbers of 
events with different combinations of ``electron--like" or ``fake--like" objects 
are then used to estimate the true composition of the CR (in terms of isolated 
electrons, non-prompt electrons from heavy-flavour decays, electrons from photon conversions 
and jets misidentified as electrons), from which the expected 
yields in the signal region can be obtained using transfer factors from the MC
simulation. 

Similar techniques are used to determine the backgrounds for the VBF--like
and VH--like categories.

\subsection{Systematic uncertainties}\label{sec:h4lsyst}
  
The dominant sources of systematic uncertainty affecting the \hzz\  8~TeV analysis
are listed in Table~\ref{tab:h4l:systs} (see Ref.~\cite{paper2012ichep} 
for the 7~TeV analysis).
\begin{table}[htbp]
  \centering
   \caption{For $\mH=125\,\gev$ and the 8~\TeV\ data analysis, the impact of the main sources 
    of systematic uncertainty specific to the \hzz\ channel on the signal yield, 
    estimated reducible background, event migration between categories and mass measurement. 
    Uncertainties common to all channels are listed in Table~\ref{tab:commonsys}.}
\vspace{0.3cm}   
\scalebox{0.68}{ % Don't ask why 0.75 is the right width for single column!!
    \begin{tabular}{lcccc}
      \hline \hline
       Source & \multicolumn{4}{c}{Uncertainty ($\%$)} \\
      \hline\hline
      \hline
      %\multicolumn{2}{l} {Signal yield \ \ \ \ \ \ \ \  $4\mu$, $2\mu2e$, $2e2\mu$, $4e$} \\
      Signal yield &  $4\mu$ & $2\mu2e$ & $2e2\mu$ & $4e$ \\
      \quad Muon reconstruction and identification       & $\pm 0.8$ & $\pm 0.4$ & $\pm 0.4$ &  - \\      
      \quad Electron reconstruction and identification   &      -    & $\pm 8.7$ & $\pm 2.4$ & $\pm 9.4$ \\
      \hline
      Reducible background (inclusive analysis)     & $\pm 24$ & $\pm 10$ & $\pm 23$ & $\pm 13$ \\
      \hline
      Migration between categories & & & & \\     
      \quad ggF/VBF/VH contributions to VBF--like cat. & \multicolumn{4}{c}{$\pm32/11/11$} \\       
      \quad $ZZ^{*}$ contribution to VBF--like cat.      & \multicolumn{4}{c}{$\pm36$} \\
      \quad ggF/VBF/VH contributions to VH--like cat.   & \multicolumn{4}{c}{$\pm15/5/6$} \\          
      \quad $ZZ^{*}$ contribution to VH--like cat.       & \multicolumn{4}{c}{$\pm30$} \\
      \hline
      Mass measurement & $4\mu$ & $2\mu2e$ & $2e2\mu$ & $4e$ \\
      \quad Lepton energy and momentum scale  & $\pm0.2$ & $\pm0.2$ & $\pm0.3$ & $\pm0.4$ \\ 
      \hline \hline
    \end{tabular}
    }
\label{tab:h4l:systs}
\end{table}
Lepton reconstruction, identification and selection efficiencies, as well as 
energy and momentum resolutions and scales, are determined using large 
control samples from the data, as described
in Section~\ref{sec:physobj}. Only the electron uncertainty contributes 
significantly to the uncertainty on the signal yield.

 The background uncertainty is dominated by the uncertainty on the transfer factors
 from the CRs to the signal region and the available number of events in the control regions. 

The uncertainty on the population of the various categories (migration) 
comes mainly from the knowledge of the theoretical cross sections for the various
production processes, the modelling of the underlying event and the 
the knowledge of the jet energy scale.

The \htollllbrief\ mass measurement is discussed in Section~\ref{mass-mu}. The main sources
contributing to the electron energy scale uncertainty are described in Section~\ref{sec:hggsyst};
 the largest impact ($\pm$0.4\%) is on the $4e$ final state. Systematic uncertainties from the
 knowledge of the muon momentum scale (discussed in detail in Ref.~\cite{new_ZZ_council2012})
are smaller.  Mass scale uncertainties related to FSR and background contamination
are below $\pm$0.1\%. 

 \begin{figure}[h!tp]
  \centering
  \includegraphics[width=0.45\textwidth]{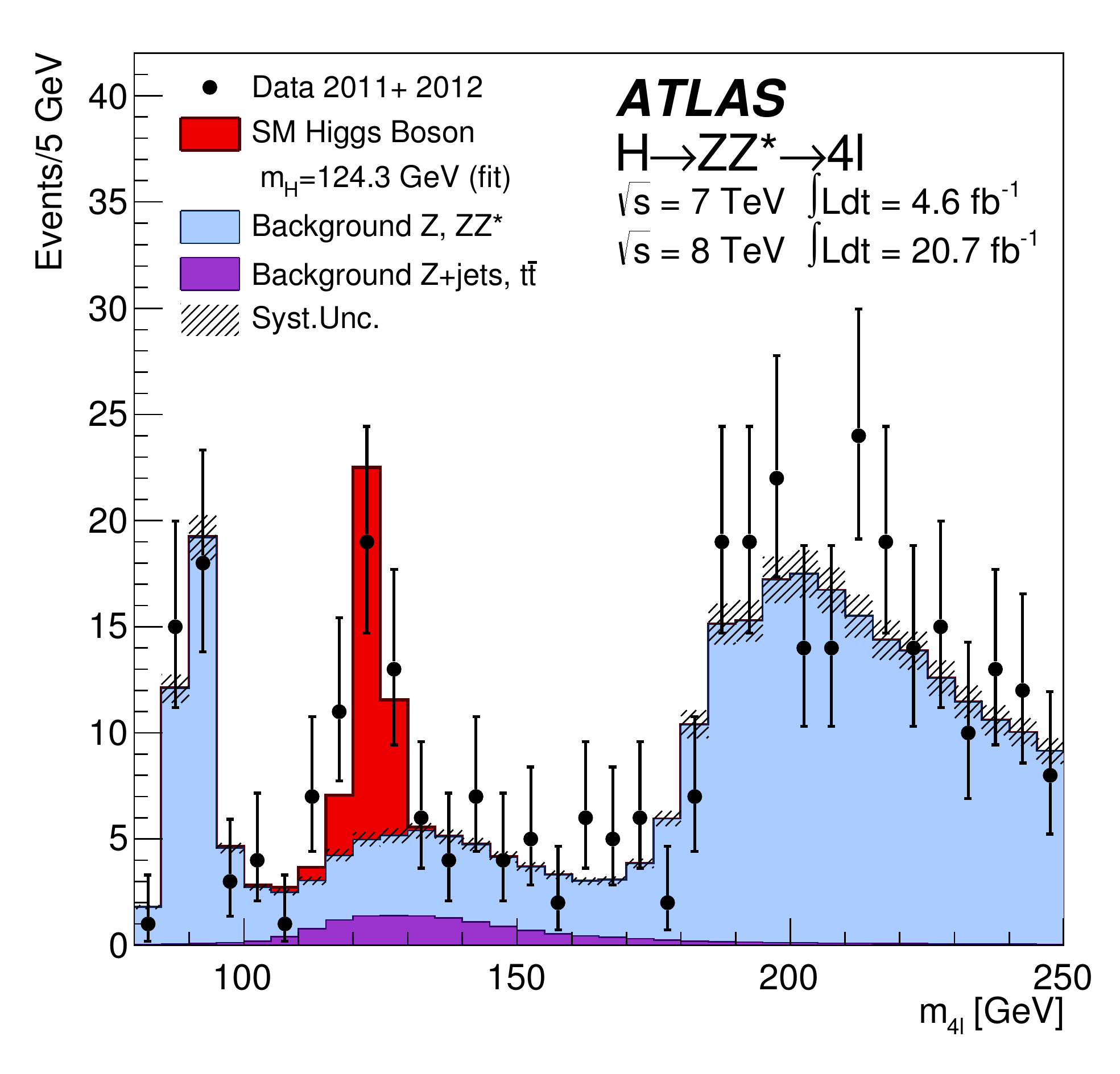}
  \vspace{-3mm}
  \caption{The distribution of the four-lepton invariant mass,
  $m_{4\ell}$, for the selected candidates in the data. 
  The estimated background, as well as the expected SM Higgs boson signal for $\mH=124.3\,\gev$  
  (scaled by the signal strength obtained from fits to the data), 
   are also shown.
  The single-resonant peak at $m_{4\ell}\sim$~90~GeV includes
  contributions from $s$-channel $Z/\gamma^*$ and $t$-channel $(Z^*/\gamma^*)(Z^*/\gamma^*)$ 
  production. \label{fig:m4lmass}}
\end{figure}

\subsection{Results}
The reconstructed four-lepton mass spectrum after all selections of the 
inclusive analysis is shown in Fig.~\ref{fig:m4lmass}.
The data are compared to the (scaled) expected Higgs 
boson signal for $\mH=124.3\,\gev$ and to the estimated backgrounds. 
 At the maximum deviation from the background-only expectation (occurring 
 at $\mH=124.3\GeV$), the significance
of the observed peak is 6.6$\sigma$ for the
combined $7\,\tev$ and $8\,\tev$ data, to be compared with 4.4$\sigma$ expected from 
SM Higgs boson production at this mass. This result
 establishes a discovery-level signal
 in the $4\ell$ channel alone. 

Table~\ref{tab:yields} presents the numbers of observed and expected events 
in the peak region. Out of a total of 32 events selected in the data, 
 one and zero candidates are found in the VBF--like and VH--like categories,
respectively, compared with an expectation of 0.7 and 0.1 events
from the signal and 0.14 and 0.04 events from the  background. 

Additional interpretation of these results is presented in Section~\ref{sec:results}.

\begin{table}[htbp]
\centering
\caption{
For the \hZZllll\ inclusive analysis, the number 
of expected signal ($\mH=125\,\gev$) and background events, 
 together with 
the number of events observed in the data, in a window of size $\pm 5\,\gev$ around 
$m_{4\ell}=125\,\gev$, 
for the combined $\sqrt{s}=7\,\tev$ and $\sqrt{s}=8\,\tev$ data.\label{tab:yields}}
\vspace{0.3cm}
\scalebox{0.9}{
\begin{tabular}{@{}ccccc@{}}
\hline\hline
                      & Signal   &      $ZZ^{*}$ & $Z+\rm{jets}$,~$t\bar{t}$ & Observed \\
\hline
    $4\mu$            & 6.3$\pm$0.8 & 2.8$\pm$0.1 & 0.55$\pm$0.15 & 13\\
%\hline
    $2e2\mu$/$2\mu2e$ & 7.0$\pm$0.6 & 3.5$\pm$0.1 & 2.11$\pm$0.37 & 13\\
%\hline
    $4e$              & 2.6$\pm$0.4 & 1.2$\pm$0.1 & 1.11$\pm$0.28 & 6\\
\hline\hline
\end{tabular}}
\end{table}

%% file: Hwwcoupl2013.tex
\input{introduction_ww}

\input{selection}

\input{control}

\input{systematics}

\input{results_ww}

%% file: introduction_ww.tex
This decay mode provides direct access to the Higgs boson couplings to $W$ bosons. 
 Its rate is large, 
  but a narrow mass peak cannot be reconstructed  
 due to the presence of two neutrinos in the final state. 
The reconstructed topology consists of two  opposite-charge leptons 
and a large momentum imbalance from the neutrinos.  
 The dominant SM backgrounds are $WW$ (which includes $WW^*$), $t\bar{t}$ and $Wt$, all of which 
 produce two $W$ bosons.  
The classification of events by jet multiplicity ($\ensuremath{N_\textrm{jet}}$)
allows the control of the background from top quarks, which contains
$b$-quark jets, as well as the extraction of the 
signal strengths for the ggF and VBF production processes.  
 For the hypothesis of a SM Higgs boson, the spin-zero initial state
and the $V\,{-}\,A$ structure of the $W$-boson decays imply a correlation 
between the directions
of the charged leptons, which can be exploited to reject the $WW$ background.
These correlations lead to the use of quantities such as the dilepton invariant mass 
$\mll$ and angular separation  $\DPhill$ in the selection criteria described below.
Drell--Yan (DY)  events ($pp\to\ZDY\to\ell\ell$) may be reconstructed 
with significant missing transverse momentum because of leptonic $\tau$ decays
or the degradation of the \met\ measurement in the high pile-up
environment of the 2012 run.  Finally, \Wjets\ production in which a jet is 
reconstructed as a lepton, and the diboson processes \Wg, $WZ$, and 
$ZZ^{\ast}$, are also significant backgrounds after all event selection.

The studies presented here are a significant update 
of those reported in Ref.~\cite{paper2012ichep}.
The signal regions considered include $ee$, $e\mu$, and $\mu\mu$ 
final states with zero,
one, or at least two reconstructed jets.
The \TwoJet\ analysis has been re-optimised to increase the sensitivity 
to Higgs boson production through VBF for $\mH=125\GeV$.
Improved DY rejection and estimation techniques have 
allowed the inclusion of $ee$ and $\mu\mu$ events from the $8\TeV$ data.
The analysis of the $7\TeV$ data, most recently documented in Ref.~\cite{ATLAS-4.7fbHWW}, 
has been updated to apply improvements from the $8\TeV$ analysis, including 
more stringent lepton isolation requirements, which reduce
the \Wjets\ background by 40\%.

%% file: selection.tex
\subsection{Event selection}\label{sec:hww-selection}

 Events are required to have two opposite-charge leptons ($e$ or $\mu$)
and to pass the same single-lepton triggers as described in Section~\ref{sec:hzz}
for the $\hzz$ channel.
The leading lepton must satisfy $\pT\,{>}\,25\GeV$ and
the sub-leading lepton  $\pT\,{>}\,15\GeV$. 
 Electron and muon identification and isolation requirements 
 (see Ref.~\cite{paper2012ichep}) are more restrictive than those
 used in the $\hzz$ analysis in order to suppress the \Wjets\ background. 

In the $\eemm$ channels, $Z{\to}\ell\ell$ and low-mass $\gamma^\ast{\to}\ell\ell$ events,
including $J/\psi$ and $\Upsilon$ production, are rejected by requiring 
$|\,\mll\,{-}\,m_Z\,|\,{>}\,15\GeV$ and  $\mll\,{>}\,12\GeV$, respectively. In the 
$\emme$ channels, low-mass $\gamma^\ast{\to}\ \tau\tau{\to}\ e\nu\nu\mu\nu\nu$ production 
is rejected by imposing
$\mll\,{>}\,10\GeV$.

Drell--Yan and multi-jet backgrounds are suppressed by requiring large
missing transverse momentum. 
For \ZeroOneJetSimple, a requirement is made on 
\mbox{$\metrel=\MET\cdot\sin|\Delta\phi_\textrm{closest}|$}, where
$\Delta\phi_\textrm{closest}$ is the smallest azimuthal angle 
between the $\vMET$ vector and any jet or high-$\pT$ charged lepton in the event;
if $|\Delta\phi_\textrm{closest}|>\pi/2$, then $\metrel=\MET$ is taken.
For additional rejection of the DY background in the \eemm\ channels with \ZeroOneJetSimple, 
the track-based \vMPT\ described in Section~\ref{sec:physobj} is used,
 modified to $\MPTRel$ in a similar way as \metrel.
For these channels, requirements are also made on \frecoil,
an estimate of the magnitude of the soft hadronic recoil 
opposite to the system consisting of the leptons
and any accompanying jet, normalised to the momentum of the system itself.
The \frecoil\ value in DY events is on average larger than that of non-DY events,
where the high-\pt\ system is balanced at least in part by recoiling neutrinos.

The \TwoJet\ analysis uses $\MET$ instead of
$\metrel$ because the larger number of jets in the final states 
reduces the signal efficiency of the $\metrel$ criterion.
For the \eemm\ channels with \TwoJet, an \met\ variant called  ``\METstvf'' is 
also employed.  In the calculation of \METstvf,
the energies of (soft) calorimeter deposits unassociated with high-$\pT$ leptons, photons, 
or jets are scaled by the
ratio of the summed scalar $\pT$ of tracks from the primary vertex unmatched 
with such objects to the summed scalar
$\pT$ of all tracks from any vertex in the event which are also unmatched with objects~\cite{atlas_metstvf}.

For all jet multiplicities, selections exploiting the
kinematic features of \hwwlnln\ events are applied.
The dilepton invariant mass is required to be small, $\mll\,{<}\,50\GeV$ for $\Njet\,{\le}\,1$
and $\mll\,{<}\,60\GeV$ for \TwoJet; the azimuthal separation of the leptons
is also required to be small, $\DPhill\,{<}\,1.8$.

%------------------------------------------------------------------------------
\subsection{Event categorisation\label{hwwcat}}
\label{sec:hww-njet}
%%%%%%%%%%%%%%%%%%%%%%%%%%%%%%%%%%%%%%%%
%          0 jets
%%%%%%%%%%%%%%%%%%%%%%%%%%%%%%%%%%%%%%%%
The analysis is divided into categories with \ZeroJet, \OneJet, and \TwoJet.
In the \ZeroJet{} analysis, $\metrel\,{>}\,25\GeV$ ($\metrel\,{>}\,45\GeV$ and $\MPTRel\,{>}\,45\GeV$) 
is required for \emme\ (\eemm) final states.  
 The transverse momentum of the dilepton system is required to be large, $\ptll\,{>}\,30\GeV$.
For \eemm\ events, the hadronic recoil is required to be typical of events with neutrinos in the final state, $\frecoil\,{<}\,0.05$.  
Finally, the azimuthal separation between the $\vpTll$ and 
$\vMET$ vectors must satisfy $\dphillMET\,{>}\,\pi/2$, in order to remove
potentially poorly reconstructed events.

%%%%%%%%%%%%%%%%%%%%%%%%%%%%%%%%%%%%%%%%%%%%%%%%%%%%%%%%%%%
%    H+1 Jet
%%%%%%%%%%%%%%%%%%%%%%%%%%%%%%%%%%%%%%%%%%%%%%%%%%%%%%%%%%
In the \OneJet{} analysis, the \metrel\ and \MPTRel\
requirements are the same as for \ZeroJet,
but the hadronic recoil threshold is looser, $\frecoil\,{<}\,0.2$.
The top-quark background is suppressed by rejecting events with a $b$-tagged jet.
The $b$-tagging algorithm described in Section~\ref{sec:physobj} is used, at an operating 
point with 85\% efficiency for $b$-quark jets and a mis-tag rate of 11\% for light-quark
and gluon jets, as measured in a sample of simulated $t\bar{t}$ events.
 The $Z\to\tau\tau$ background in \emme\ final states is suppressed using 
an invariant mass $m_{\tau\tau}$ computed assuming
that the neutrinos from $\tau$ decays are collinear with the charged
leptons~\cite{CollApp} and that they are
the only source of \MET. The requirement $|\,m_{\tau\tau}\,{-}\,m_{Z}\,|\,{\ge}\,25\GeV$
is applied.

The \TwoJet\ analysis is optimised for the selection of the VBF production process.
The two leading jets, referred to as ``tagging jets'', are required to have a large rapidity 
separation, $\Dyjj\,{>}\,2.8$, and a high invariant mass, $m_{jj}\,{>}\,500\GeV$. 
To reduce the contribution from ggF, events containing any
jet with $\pT\,{>}\,20\GeV$ in the rapidity gap between the two tagging jets are rejected. 
Both leptons are required to be in the rapidity gap.
The DY background is suppressed by imposing $\MET\,{>}\,20\GeV$ for \emme, and
$\MET\,{>}\,45\GeV$ and $\METstvf\,{>}\,35\GeV$ for \eemm.  
The same $Z\to\tau\tau$ veto and $b$-jet veto are applied as in the \OneJet\ analysis.
The $t\bar{t}$ background is further reduced by requiring a small
total transverse momentum, $|\pTtot|\,{<}\,45\GeV$, where
$\vpTtot\,{=}\,\vpTll\,{+}\,\pTvec^{\rm jets}\,{+}\,\vMET$, and $\pTvec^{\rm jets}$ is 
the vectorial sum of all jets in the event with $\pT\,{>}\,25\GeV$.

 The total signal selection efficiency for $\hwwlnln$ events produced with
 $\ell=e,\mu$, including all the final state topologies considered,
 is about 5.3\% at 8 TeV for a Higgs boson mass of 125 GeV. 

The dilepton transverse mass $\mT$ is the discriminating variable used in the 
fit to the data to extract the signal strength.  
 It is defined as
$\mT\,{=}\,((E_{\rm T}^{\ell\ell}\,{+}\,\MET)^{2}\,{-}\,|\,\vpTll\,{+}\,\vMET\,|^{2})^{1/2}$
with $E_{\rm T}^{\ell\ell}\,{=}\,(|\,\vpTll\,|^{2}\,{+}\,\mll^{2})^{1/2}$.  For the 
\emme\ channels with $\Njet\,{\le}\,1$, the fit is performed
separately for events with  $10\GeV\,{<}\,\mll\,{<}\,30\GeV$ and events with $30\GeV\,{<}\,\mll\,{<}\,50\GeV$, since
 the signal-to-background ratio varies across the  $\mll$ distribution, as shown
 in Fig.~\ref{fig:hww-Mll}.

%% file: control.tex
\subsection{Background estimation}
\label{sec:hww-control}

The leading SM processes producing two isolated high-$\pt$ leptons 
and large values of $\MET$ 
are $WW$ and top-quark production, where the latter 
 includes (here and in the following) both $t\bar{t}$ 
and single top-quark processes ($tW$, $tb$ and $tqb$). 
These backgrounds, as well as $Z\to\tau\tau$, are normalised to the data in
control regions defined by selections similar to those used for the signal region,
but with some criteria reversed or modified to obtain signal-depleted samples enriched
in particular backgrounds.  
The event yield in the CR (after subtracting contributions from processes other 
than the targeted one) is extrapolated to the signal region using transfer factors
obtained from MC simulation.

Additional significant backgrounds arise from \Wjets\ and $\ZDY$, which are
dissimilar to the signal but have large cross sections.  A small fraction of these pass 
the event selection through rare final-state configurations and/or mis-measurements.
This type of background is difficult to model reliably with the simulation and is therefore 
estimated mainly from data.

A third category of background consists of diboson processes with smaller cross 
 sections, including \Wg, $WZ$, and $ZZ^{\ast}$ (inclusively indicated in the 
 following as {\em Other $VV$}),
 and the $WW$ background in the \TwoJet\ analysis.  These processes are estimated using 
the MC simulation normalised to the NLO cross sections from MCFM~\cite{Campbell:1999ah}, except
for the \TwoJet\ $WW$ background, for which the cross section from 
the relevant MC generators (see Table~\ref{tab:gen}) is used.
The {\em Other $VV$} processes all produce same-charge and opposite-charge lepton pairs, as 
does \Wjets.  The number and kinematic features of same-charge events 
which would otherwise pass the 
full event selection are compared to the above-mentioned predictions 
for these backgrounds,
 and good agreement is observed.

\subsubsection{\Wjets}
\label{sec:hww-wjets-control}

The \Wjets{} background is estimated using a CR in the data
in which one of the two leptons satisfies the identification and
isolation criteria, and the other	
lepton (denoted here as ``anti-identified'') fails these criteria 
but satisfies looser requirements. All other analysis selections are applied.
The contribution to the signal region is then obtained by scaling the number of 
events in the CR by transfer factors,  
defined as the ratio of the number of fully identified lepton 
candidates passing all selections to the number of anti-identified leptons.  
The transfer factors are obtained from a dijet sample as a function of the $\pt$ and $\eta$ 
of the anti-identified lepton.

\subsubsection{\ZDY}
\label{sec:hww-zll-control}

The $\ZDY$ yield in the {\eemm} channels for $\Njet\,{\le}\,1$ is estimated using the 
$\frecoil$ requirement efficiency in data for DY and non-DY processes.  
The former is measured in  \eemm\  events in the $Z$-boson peak region.  
The latter is measured in the \emme\ signal region,
taking advantage of the fact that the $\frecoil$ distribution is nearly identical for all 
non-DY processes including the signal, as well as for \emme\ and \eemm\ final states.
The DY normalisation in 
the \eemm\ signal region can then be extracted, given the two measured efficiencies 
and the total number of events in the \eemm\ signal region before and after 
the \frecoil requirement.
For the {\eemm} channels with \TwoJet, the two-dimensional distribution 
($\MET$, $\mll$) in the data is used to estimate the total $\ZDY$ yield, 
as in Ref.~\cite{ATLAS-4.7fbHWW}.

The $Z\to\tau\tau$ background is normalised to the data using an \emme\ CR 
defined by the back-to-back configuration of the leptons, $\DPhill\,{>}\,2.8$.
For the corresponding CR with \TwoJet, no $b$-tagged jets are allowed, and $|\pTtot|\,{<}\,45\GeV$ is required
in addition, in order to reduce the contamination from top-quark production. 
A separate CR in the $\Zll$ peak region is used to correct the modelling
of the VBF-related event selection.

\subsubsection{$t\bar{t}$ and single top-quark}
\label{sec:top-control}
The top-quark background for the \ZeroJet\ category is estimated using the 
procedure described in Ref.~\cite{paper2012ichep}, namely 
from the number of events in data with any number of reconstructed
jets passing the \metrel\ requirement (a sample dominated by top-quark production), 
 multiplied by the fraction of top-quark
events with no reconstructed jets obtained from simulation.  This estimate is corrected
using a CR containing $b$-tagged jets.
The top-quark background in the \OneTwoJetSimple\ channels is normalised to the data in a
CR defined by requiring exactly one $b$-tagged jet and all other signal selections 
except for the requirements on $\DPhill$ and $\mll$.

\subsubsection{$WW$}
\label{sec:hww-WW-control}

\begin{figure}[t!]
  \centering
 \hspace*{-0.3cm}\includegraphics[width=.52\textwidth]{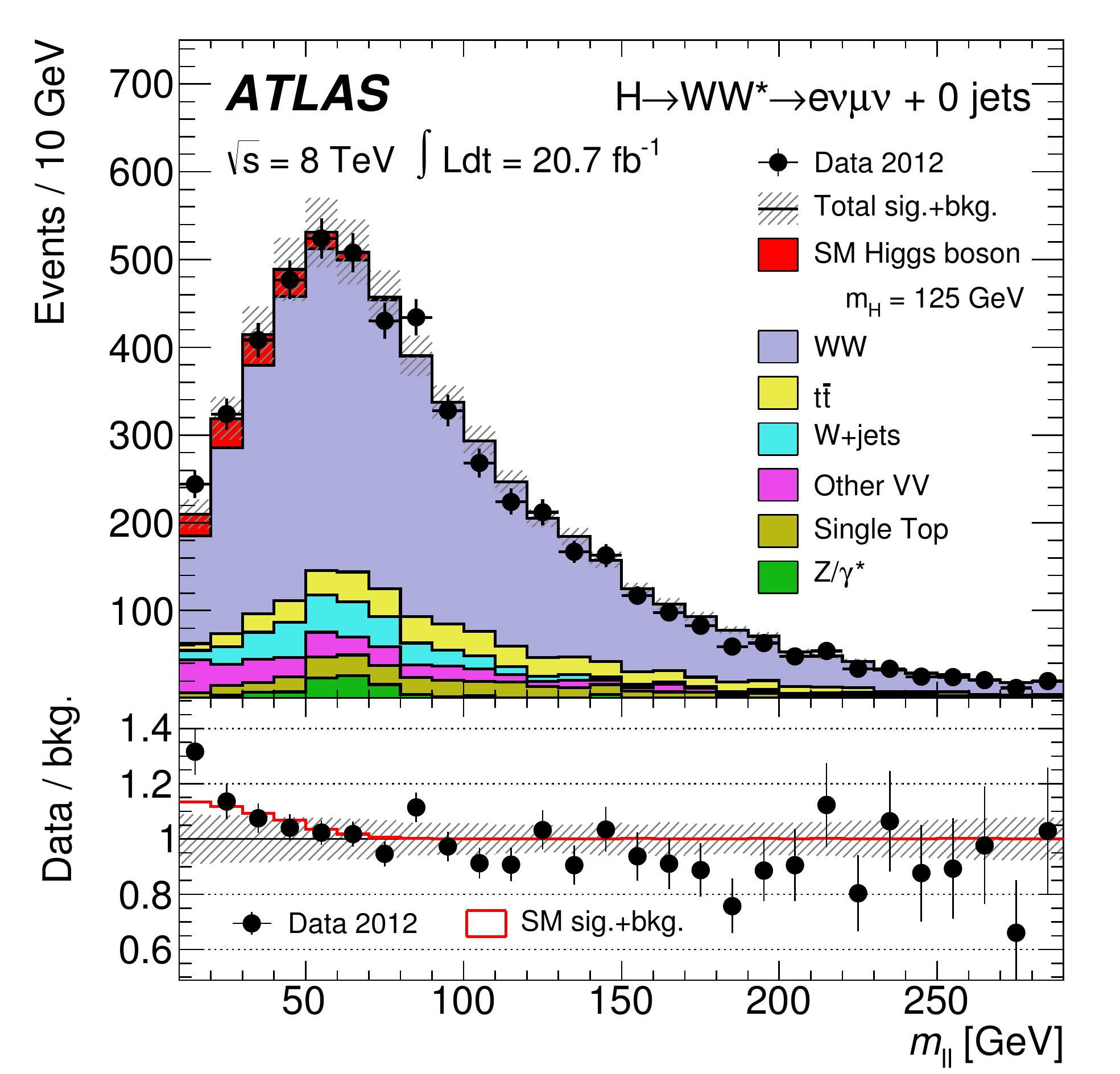}
  \caption{The $\mll$ distribution of \emme\ events with \ZeroJet\ for the 8~TeV \hwwlnln\ analysis.
    The events with $\mll < 50\GeV$ correspond to the signal region except that the $\DPhill < 1.8$ requirement
    is not applied here, and the events with $50\GeV < \mll < 100\GeV$ correspond to the \ZeroJet\ $WW$ control
    region.      The signal is stacked on top of the background.
       The hatched area represents the total uncertainty on the sum of the signal and
    background yields from statistical, experimental, and theoretical sources.
    The lower part of the figure shows the ratio of the data to the predicted background.  For comparison,
	the expected ratio of the signal plus background to the background alone is also shown.
  }
  \label{fig:hww-Mll}
\end{figure}

The $WW$ background for $\Njet\,{\le}\,1$ is normalised using CRs in data 
defined with the same selection as the signal region except that the
$\DPhill$ requirement is removed and the \mll\ bound is modified:  for \ZeroJet\  
$50\GeV\,{\le}\,{\mll}\,{<}\,100\GeV$ is required,
 while for \OneJet\ $\mll\,{>}\,80\GeV$ is used to define the CR.
 Figure~\ref{fig:hww-Mll} shows the $\mll$ distribution of \emme\ events with \ZeroJet\ in the 8 TeV data. 
 The level of agreement between the predicted background and the data for $\mll > 100\GeV$, a region with 
 negligible signal contribution, validates the $WW$ background normalisation and the 
 extrapolation procedure based on the simulation.
The \TwoJet\ prediction is taken from
simulation because of the difficulty of isolating a kinematic region with enough
events and small contamination from the top-quark background.

%% file: systematics.tex
\subsection{Systematic uncertainties}
\label{sec:hww-systematics}

 The systematic uncertainties affecting this analysis
are summarized here and described in detail in Ref.~\cite{new_WW_Moriond2013}. 
The leading sources, i.e., those resulting in at least 4\% uncertainty on the total signal or 
background yield in at least one $\Njet$ category, are reported in Table~\ref{hww_systematics}.
\begin{table}[t!]
\centering
\caption{ 
For $m_H=125$~GeV, the leading systematic uncertainties on the total signal and background yields 
  for the $8\TeV$ \hwwlnln\ analysis.  All numbers are summed 
over lepton flavours. Sources contributing less than 4\% are omitted, and individual entries 
below 1\%
are indicated with a '-'.
% The QCD scale uncertainties on the ggF signal process are anti-correlated between the exclusive $\Njet$ modes. 
Relative signs indicate correlation and anticorrelation (migration) between the $\Njet$ categories represented by 
adjacent columns, and a $\pm$ indicates an uncorrelated uncertainty.  The exception is the jet energy
scale and resolution, which includes multiple sources of uncertainty treated as correlated across categories 
but uncorrelated with each other.  All rows are uncorrelated.
}
\label{hww_systematics}
\vspace{0.3cm}
\scalebox{0.75}{
  \begin{tabular}{lrrr}
    \dbline
    Source & \ZeroJet\ & \OneJet\  & \TwoJet\  \\ % & \ZeroJet\ & \OneJet\  & \TwoJet\  \\
    \sgline
\multicolumn{4}{l}{Theoretical uncertainties on total signal yield (\%)} \\
\quad QCD scale for ggF, $\Njet\,{\ge}\,0$	& $+$13 &      - &       -  \\	% &   -  &   -  &   -	\\
\quad QCD scale for ggF, $\Njet\,{\ge}\,1$	& $+$10 &  $-$27 &       -  \\	% &   -  &   -  &   -	\\
\quad QCD scale for ggF, $\Njet\,{\ge}\,2$	&   - 	&  $-$15 &    $+$4  \\	% &   -  &   -  &   -  	\\
\quad QCD scale for ggF, $\Njet\,{\ge}\,3$	&   -   &     -  &    $+$4  \\	% &   -  &   -  &   -  	\\
\quad Parton shower and underlying event      	&  $+$3 &  $-$10 &  $\pm$5  \\	% &   -  &   -  &   -  	\\
%                             \quad PDF     &   8 &   7 &   3  \\	% &   1  &   1  &   1 	\\
                \quad QCD scale (acceptance)    &  $+$4 &   $+$4 &  $\pm$3  \\	% &   -  &   -  &   - 	\\
    \sgline
\multicolumn{4}{l}{Experimental uncertainties on total signal yield (\%)} \\
       \quad Jet energy scale and resolution    &      5 &      2 &      6	\\	% &   2  &   3  &   7  	\\
    \sgline
\multicolumn{4}{l}{Uncertainties on total background yield (\%)} \\
       \quad Jet energy scale and resolution    &      2  &       3  &       7  	\\
          \quad $WW$ transfer factors (theory)  & $\pm$1  &  $\pm$2  &  $\pm$4  	\\
                \quad $b$-tagging efficiency    &      -  &    $+$7  &    $+$2 		\\
                 \quad $\frecoil$ efficiency    & $\pm$4  &  $\pm$2  &   - 	\\
    \dbline
  \end{tabular}
}
\end{table}

 Theoretical uncertainties on the 
inclusive signal production cross sections are given in 
Section~\ref{sec:physobj}.  Additional, larger uncertainties from % missing higher-order corrections 
the QCD renormalisation and factorisation scales 
affect the predicted distribution of the ggF signal among the exclusive jet bins 
and can produce migration between categories.  These uncertainties
are estimated using the HNNLO program~\cite{Catani:2007vq,Grazzini:2008tf}
and the method reported in Ref.~\cite{Stewart:2011cf}. Their impact 
on the signal yield is summarised in Table~\ref{hww_systematics}, in addition to
other non-negligible contributions (parton shower and underlying event modelling, as well as
 acceptance uncertainties due to QCD scale variations).

The experimental uncertainties affecting the expected signal and background yields
are associated primarily with the reconstruction and identification efficiency,
 and with the energy and momentum
scale and resolution, of the final-state objects (leptons, jets, and \met),
as described in Section~\ref{sec:physobj}. The largest impact 
on the signal expectation
comes from the knowledge of the 
jet energy scale and resolution (up to 6\% in the \TwoJet\ channel). 

For the backgrounds normalised using control regions, uncertainties come from
the numbers of events in the CR and the contributions of other processes, as well
as the transfer factors to the signal region.

 For the $WW$ background in the \ZeroOneJetSimple\ final states, 
  the theoretical uncertainties on the transfer factors
 (evaluated according to the prescription of 
 Ref.~\cite{LHCHiggsCrossSectionWorkingGroup:2012vm})
 include the impact of missing higher-order QCD corrections, 
 PDF variations, and MC modelling choices. They  amount to 
 $\pm$2\% and $\pm$4--6\% relative to the predicted $WW$ background 
 in the \ZeroJet\ and \OneJet\ final states, respectively. 
 For the $WW$ yield in the \TwoJet\ channel, which is obtained from simulation, 
  the total uncertainty is $42\%$ for QCD production with gluon emission, and 11\% for
 the smaller but non-negligible contribution from purely electroweak processes; the latter 
 includes the size of possible interference with Higgs boson 
 production through VBF. The resulting uncertainties on the total
  background yield for all $\Njet$ are quoted in Table~\ref{hww_systematics}.
  % and reflect the dominance of this background, particularly for \ZeroOneJet.

 The leading uncertainties on the top-quark background are experimental. The $b$-tagging 
 efficiency is 
 the most important of these, and it appears in Table~\ref{hww_systematics} primarily through its 
 effect on this background.  Theoretical uncertainties on the top-quark background have the 
 greatest relative importance, $\pm$2\% on the total background yield, for \TwoJet,
 and therefore do not appear in Table~\ref{hww_systematics}. 

% The dominant uncertainties on the top background for the \ZeroJet\ analysis, about 10\%,  
% are related to the MC-based estimation of the number of events with zero jets.
% For the \TwoJet\ case, where a CR is used, an additional $15\%$ modelling uncertainty is evaluated by
% comparing the transfer factors between various generators.
%The uncertainty on the $b$-jet tagging efficiency calibration is the primary
%uncertainty on the transfer factor for the top background for \OneTwoJet.
The \Wjets\ transfer factor uncertainty ($\pm$(40--45)\%) is dominated by differences in the jet 
composition between dijet and \Wjets\ samples as observed in the MC simulation. 
The uncertainties on the muon and electron transfer factors are treated as correlated
among the $\Njet$ categories but uncorrelated with each other.
 The impact on the total background uncertainty is at most $\pm$2.5\%. 
  The main uncertainty on the DY contribution in the \ZeroOneJetSimple\ channels 
  comes from the use of the \frecoil\ efficiency evaluated at the peak of the $Z$-boson
  mass distribution for 
  the estimation of the DY contamination in the low-$m_{\ell\ell}$ region.  
% yielding a total uncertainty of
% \ZeemmSRErrorZerojet\ and \ZeemmSRErrorOnejet\ in the \ZeroOneJet\ modes, respectively.

% Lepton momentum scale uncertainties are also propagated to the $\MPT$ calculation.
% In addition, uncertainties are assigned to the scale and resolution
% of the remaining $\MPT$ component not associated with
% charged leptons. These uncertainties are calculated by comparing the properties of $\MPT$ in $Z$
% events in real and simulated data, as a function of the sum of the hard {\pT} objects in the event.

   The uncertainty on the \mT\ shape for the total background, which is used in the fit
   to extract the signal yield, is dominated by the
   uncertainties on the normalisations of the individual components. 
   The only explicit \mT\ shape uncertainty is applied to the $WW$ background, and is 
   determined by comparing several generators and showering algorithms. 
   
 The estimated background contributions with their uncertainties are listed in
 Table~\ref{hww_comparison}.

% %------------------------------------------------------------------------------
% \subsubsection{Uncertainties on backgrounds normalised to control regions}
% 
% For the backgrounds normalised using 
% CRs ($WW$ for the \ZeroJet\ and \OneJet\ analyses and top in the \OneTwoJet\ analyses), 
% the sources of uncertainty can be grouped into four categories:
% the statistical uncertainty, 
% the theoretical and experimental uncertainties on the simulation-based extrapolation from the CR
% to the signal region, and the uncertainty on the other contributing processes in the
% CR, which are subtracted from the data yield to get the estimated number
% of events from the targeted background.  

%% file: results_ww.tex
\subsection{Results}
\label{sec:hww-results}
\begin{figure}[t!]
  \centering
\subfigure[]{\label{fig:hww-MT01jet}\hspace*{-0.5cm}\includegraphics[width=.5\textwidth]{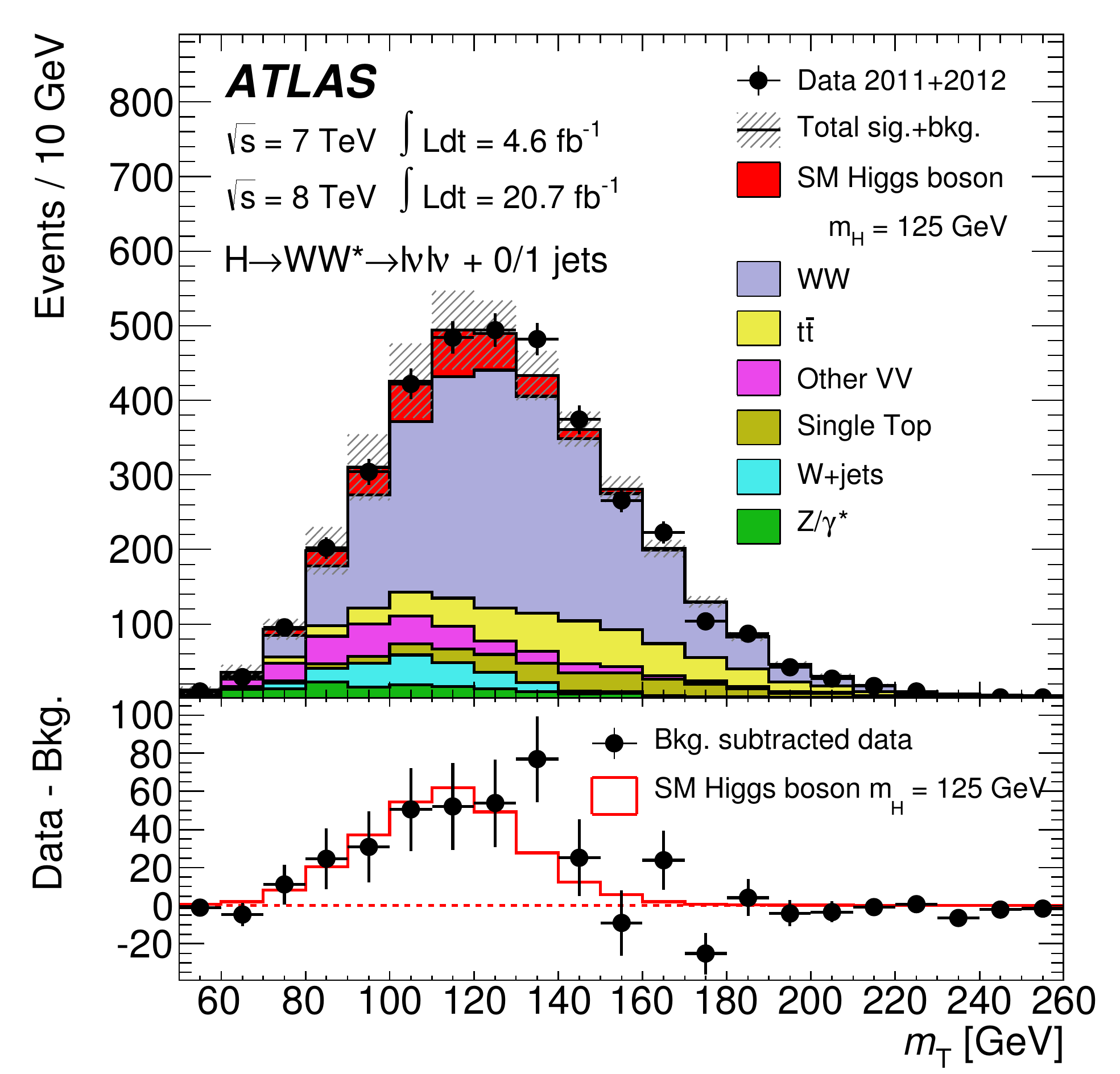}}
\subfigure[]{\label{fig:hww-MT2jet}\hspace*{-0.5cm}\includegraphics[width=.5\textwidth]{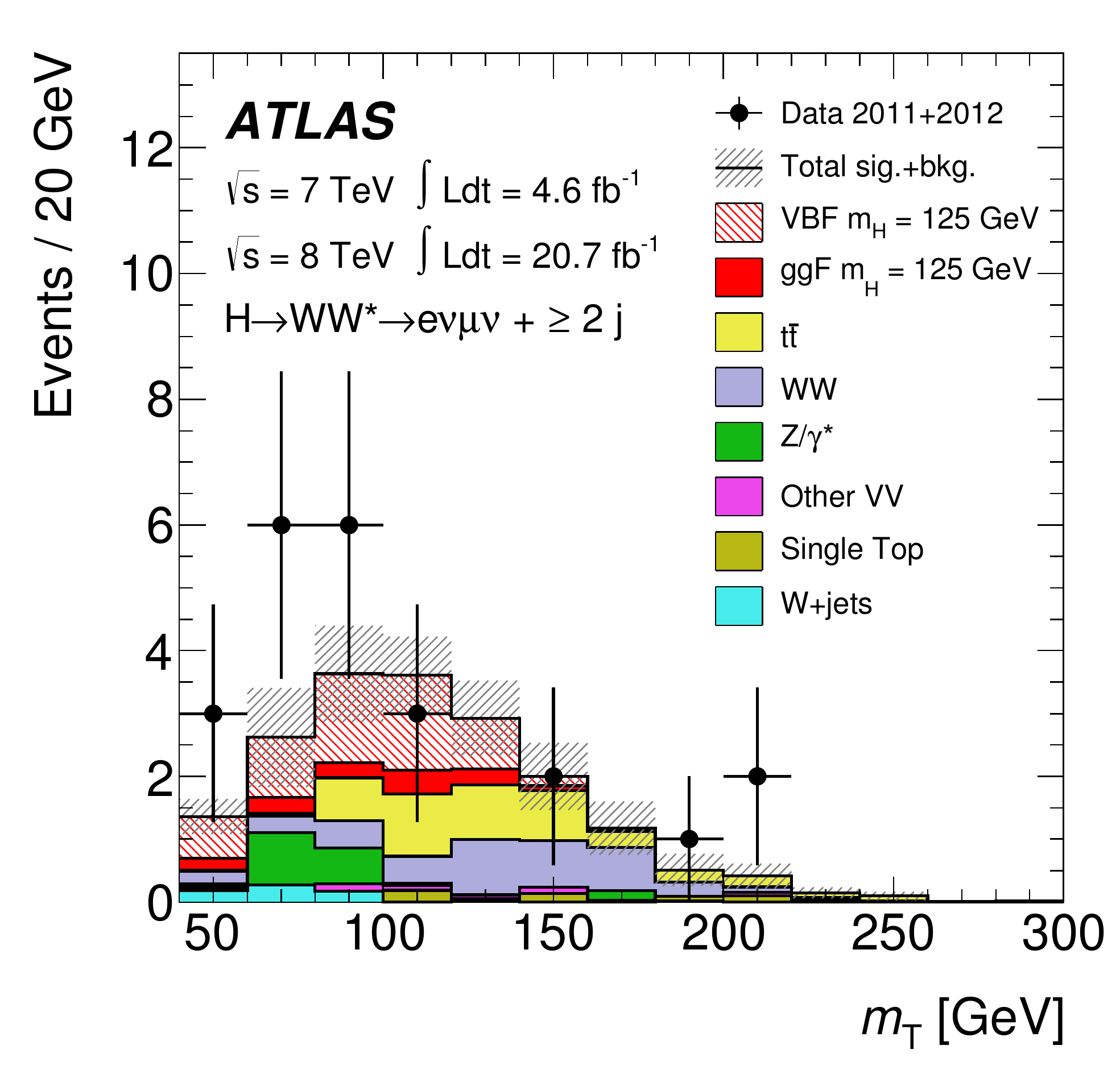}}
  \caption{The transverse mass distributions for events passing the full selection of the \hwwlnln\ analysis:
      (a) summed over all lepton flavours for final states with \ZeroOneJetSimple; (b) different-flavour
      final states with \TwoJet.
     The signal is stacked on top of the background, and in (b) is shown separately 
     for the ggF and VBF production processes. The hatched area represents the total uncertainty 
     on the sum of the signal and
    background yields from statistical, experimental, and theoretical sources.
       In the lower part of (a), the residuals of the data with respect to the estimated background
	 are shown, compared to the expected $\mT$ distribution of a SM Higgs boson.
  }
  \label{fig:hww-MT}
\end{figure}

Figure~\ref{fig:hww-MT} shows the transverse mass distributions after the full selection
for \ZeroOneJetSimple\ and \TwoJet\ final states. The regions with $\mT\,{>}\,150\GeV$ 
are depleted of signal contribution;  the level of agreement 
 of the data with the expectation in these regions, which are different from those used to normalise the
   backgrounds, illustrates the quality of the background estimates. 
The expected numbers of signal and background events at $8\TeV$ 
are presented in Table~\ref{hww_comparison}. 
The VBF process contributes 2\%, 12\% and 81\% of the predicted signal 
in the \AllJet\ final states, respectively.  
The total number of observed events in the same $\mT$ windows as in Table~\ref{hww_comparison} 
is 218 in the $7\TeV$ data and 1195 in the $8\TeV$ data.

An excess of events relative to the background-only expectation is observed in the data, with 
the maximum deviation (\MinpzerosigmaComb $\sigma$) occuring at $\mH\,{=}\,\MinpzeromassComb\GeV$. 
 For $\mH\,{=}\,125.5\GeV$, a significance of \NompzerosigmaExpComb $\sigma$ is observed,
  compared with an expected value of \NompzerosigmaExpComb $\sigma$ for a SM Higgs boson.
 
Additional interpretation of these results is presented in Section~\ref{sec:results}.

\begin{table}[tb!]
\caption{
For the \hwwlnln\ analysis of the $8\TeV$ data, the numbers of events observed in the
data and expected from signal ($\mH\,{=}\,125.5\GeV$) and backgrounds inside the transverse mass regions
$0.75\,m_{H}\,{<}\,\mT\,{<}\,m_{H}$ for $\Njet\,{\le}\,1$ and $\mT\,{<}\,1.2\,m_{H}$ for \TwoJet.
 All lepton flavours are combined. The total background as well as its main components
 are shown.  The quoted uncertainties include the statistical and systematic contributions, and 
 account for anticorrelations between the background predictions.}
\label{hww_comparison}
\vspace{0.3cm}
%\resizebox{\textwidth}{!}{
\scalebox{0.9}{
\begin{tabular}{lr@{$\,\pm \,$}l r@{$\,\pm \,$}l r@{$\,\pm \,$}l}
\dbline
%   & \multicolumn{2}{c}{Signal} & \multicolumn{2}{c}{$WW$} &  \multicolumn{2}{c}{$WZ/ZZ/W\gamma$} & \multicolumn{2}{c}{$t\bar{t}$} & \multicolumn{2}{c}{$tW/tb/tqb$} &\multicolumn{2}{c}{$Z/\gamma^{\ast}+\mathrm{jets}$} &  \multicolumn{2}{c}{$W+\mathrm{jets}$} &   \multicolumn{2}{c}{Total Bkg.} & Obs. \\
  & \multicolumn{2}{c}{\ZeroJet} & \multicolumn{2}{c}{\OneJet} &  \multicolumn{2}{c}{\TwoJet}  \\
\sgline
 Observed  & \multicolumn{2}{c}{831} & \multicolumn{2}{c}{309} &  \multicolumn{2}{c}{55}  \\
Signal & 100  & 21  & 41   & 14   & 10.9 & 1.4  \\
 Total background & 739 & 39 & 261 & 28 & 36  & 4  \\
\sgline
 $WW$ 	& 551 & 41  & 108 & 40  &  4.1 & 1.5 	\\
 {\em Other $VV$} &  58 & 8   &  27 &  6  &  1.9 & 0.4 	\\
 Top-quark	&  39 & 5   &  95 & 28  &  5.4 & 2.1 	\\
 \Zjets &  30 & 10  &  12 &  6  & 22   & 3  	\\
 \Wjets &  61 & 21  &  20 &  5  &  0.7 & 0.2 	\\
 \dbline
\end{tabular}
}
\end{table}

%% file: Methods.tex
The statistical treatment of the data is described in
Refs.~\cite{paper2012prd,LHC-HCG,Moneta:2010pm,HistFactory,ROOFIT}.
 Hypothesis testing and confidence intervals are based on the profile
likelihood ratio~\cite{Cowan:2010st} $\Lambda(\vec\alpha)$. The latter depends on 
one or more parameters of interest $\vec\alpha$, 
such as the Higgs boson production strength $\mu$ normalised to the 
SM expectation (so that $\mu=1$ corresponds to the SM Higgs boson
hypothesis and $\mu=0$ to the background-only hypothesis), mass $m_H$, 
coupling strengths $\vec\Cc$,  ratios
of coupling strengths $\vec\Rr$, as well as on nuisance parameters $\vec\theta$:

\begin{equation}
  \Lambda(\vec\alpha) = \frac{L\big(\vec\alpha\,,\,\hat{\hat{\vec\theta}}(\vec\alpha)\big)}
                              {L(\hat{\vec\alpha},\hat{\vec\theta})\label{eq:LH}} 			     
\end{equation}

The likelihood functions in the numerator and denominator of the above equation
are built using sums of signal and background probability density functions (pdfs) in the discriminating variables
 (chosen to be the $\gamma\gamma$ and $4\ell$ mass spectra for \hgg\ and \hZZllll, respectively, 
and the $\mT$ distribution for the \hWWlvlv\ channel). The pdfs are derived from MC simulation 
for the signal and from both data and simulation for the background, as  described in 
Sections~\ref{sec:hgg}--\ref{sec:hww}. Likelihood fits to the observed data are done for the
parameters of interest. 
% The likelihood functions in the numerator and denominator of the above equation
% are built on sums of signal and background probability density 
% functions, derived from MC simulation for the signal and from
% both data and simulation for the background, as described in 
% Sections~\ref{sec:hgg}--\ref{sec:hww}. For each chosen set of values for the parameters $\vec\alpha$, 
% the likelihood functions are fit to the distributions of 
% discriminating variables in the data (chosen to 
% be the $\gamma\gamma$ and $4\ell$ mass spectra for \hgg\ and \hZZllll, respectively, 
% and the $\mT$ distribution for the \hWWlvlv\ channel). 
 The single circumflex in Eq.~(\ref{eq:LH}) denotes the unconditional maximum
likelihood estimate of a parameter and the double circumflex 
denotes the conditional maximum likelihood estimate for given
fixed values of the parameters of interest $\vec\alpha$. 
 Systematic uncertainties and their correlations~\cite{paper2012prd}
are modelled by introducing nuisance parameters $\vec\theta$
 described by likelihood functions associated with the estimate
 of the corresponding effect. The choice of the
parameters of interest depends on the test under consideration, with 
the remaining parameters being ``profiled", {\it i.e.}, similarly to nuisance 
parameters they are set to the values that maximise the likelihood 
function for the given fixed values of the parameters of interest.

%% file: Mass-mu.tex
The mass of the new particle is measured from the data using
the two channels with the best mass resolution, 
 $H\to\gamma\gamma$ and \hZZllll. In the two cases,  
 $m_H=126.8 \pm 0.2\,\text{(stat)}\, \pm 0.7 \,\text{(sys)}\,\GeV$ 
 and
 $m_H=124.3^{+0.6}_{-0.5}\,\text{(stat)}\,^{+0.5}_{-0.3}\,\text{(sys)}\, \GeV$ 
 are obtained from fits to the mass spectra.  

 To derive a combined mass measurement, 
 the profile likelihood ratio $\Lambda(m_H)$ is used; the 
 signal production strengths $\mu^{\gamma\gamma}$ and $\mu^{4\ell}$, 
 giving the signal yields measured in the two individual channels normalised to the SM expectation,
 are treated as independent nuisance parameters in order to allow for the possibility
 of different deviations from the SM prediction in the two decays modes. 
The ratios of the cross sections for the various production modes for each 
channel are fixed to the SM values. It was verified that this restriction does not 
cause any bias in the results. The combined mass is measured to be:

\begin{equation}
m_H = 125.5 \pm 0.2\,\text{(stat)}\, ^{+0.5}_{-0.6}\,\text{(sys)}\, \GeV \label{eq:mass}
\end{equation}

As discussed in Sections~\ref{sec:hggsyst} and \ref{sec:h4lsyst}, the main sources 
of systematic uncertainty are the photon and lepton energy and momentum scales. 
 In the combination, the consistency between the muon and electron final states 
in the \hZZllll\ channel causes a $\sim0.8\sigma$ adjustment of the 
overall $e/\gamma$ energy scale, which translates into a $\sim350$\MeV\ downward 
shift of the fitted $m_{H}^{\gamma\gamma}$ value with respect to 
the value measured from the \hgg\ channel alone.

To quantify the consistency between the fitted $m_{H}^{\gamma\gamma}$ and $m_H^{4\ell}$ 
masses, the data are fitted with the profile likelihood ratio $\Lambda(\Delta m_H)$, 
where the parameter of interest is the mass difference 
$\Delta m_{H} = m_{H}^{\gamma\gamma} - m_H^{4\ell}$. The
average mass $m_H$ and the signal strengths $\mu^{\gamma\gamma}$ and $\mu^{4\ell}$ 
are treated as independent nuisance parameters.
The result is:

\begin{equation}
\Delta m_{H} =              2.3 ^{+0.6}_{-0.7}\,\text{(stat)}\, \pm 0.6\,\text{(sys)}\, \GeV
\end{equation}

\noindent where the uncertainties are 68$\%$ confidence intervals 
computed with the asymptotic approximation~\cite{Cowan:2010st}.
 From the value of the likelihood  
 at $\Delta m_{H}=0$, the probability for a single
Higgs boson to give
a value of $\Lambda(\Delta m_H)$ disfavouring the $\Delta
m_{H} = 0$ hypothesis more strongly than observed in the data
is found to be at the level of 1.2\% ($2.5\sigma$) 
using the asymptotic approximation,
and 1.5\% ($2.4\sigma$) using Monte Carlo ensemble tests. 
 In order to test the effect of a possible non-Gaussian behaviour of 
 the three principal sources contributing to the 
 electron and photon energy scale systematic uncertainty
(the \Zee\ calibration procedure, the knowledge of the material upstream 
of the electromagnetic calorimeter and the energy scale of the presampler detector)
 the consistency between the two mass measurements is also evaluated 
 by considering $\pm1\sigma$ values for these uncertainties. 
 With this treatment, the consistency increases to up to 8\%.

%%%%%%%%%%%%%%%%%%%%%%%%%%%%%%%%%%%%%%%%%%%%%%%%%%%%%%%%%%%%%
%%%%%%%%%%%%%%%%%%%%%%%%%%%%%%%%%%%%%%%%%%%%%%%%%%%%%%%%%%%%%
%%%%%%%%%%%%%%%%%%%%%%%%%%%%%%%%%%%%%%%%%%%%%%%%%%%%%%%%%%%%%
% SIGNAL STRENGTH
%%%%%%%%%%%%%%%%%%%%%%%%%%%%%%%%%%%%%%%%%%%%%%%%%%%%%%%%%%%%%
%%%%%%%%%%%%%%%%%%%%%%%%%%%%%%%%%%%%%%%%%%%%%%%%%%%%%%%%%%%%%

 To measure the Higgs boson production strength, the parameter $\mu$ is determined from a fit to 
the data using the profile likelihood ratio $\Lambda(\mu)$ for a 
fixed mass hypothesis corresponding to the measured value $m_H = 125.5$~\GeV.
The results are shown in Fig.~\ref{fig:muhat_overview_WWZZgg_long}, where  
 the production strengths measured in the three channels and in their main analysis categories
  are presented. The signal production strength normalised to the SM expectation, obtained
  by combining the three channels, is: 
 
\begin{equation}
\mu = 1.33 \pm 0.14\,\text{(stat)}\pm 0.15\,\text{(sys)}\label{eq:mu}
\end{equation} 

\noindent where the systematic uncertainty receives similar contributions from 
the theoretical uncertainty on the
signal cross section (ggF QCD scale and PDF, see Table~\ref{tab:commonsys}) and 
all other, mainly experimental, sources. The uncertainty on the mass 
measurement reported in Eq.~(\ref{eq:mass}) produces a $\pm3$\% variation of $\mu$.  
 The consistency between this measurement and the SM Higgs 
boson expectation ($\mu=1$) is about 7\%; the use of a flat likelihood for
the ggF QCD scale systematic uncertainty in the quoted $\pm1\sigma$ interval
yields a similar level of consistency with the $\mu=1$ hypothesis. 
 The overall compatibility between the signal strengths measured in the three final states 
and the SM predictions is about 14$\%$, with the largest deviation 
($\sim1.9\sigma$) observed in the \hgg\ channel.  
 Good consistency between the
measured and expected signal strengths is also found for the various categories of 
the \hgg, \hZZllll\ and \hWWlnln\ analyses, 
 which are the primary experimental inputs to the fit discussed in this section.   
 If the preliminary $H\to\tau\tau$~\cite{tautauHCP} 
and \hbb~\cite{new_bb_hcp2012} results,
 for which only part of the 8~TeV dataset is used (13\,\ifb),
 were included, the combined signal strength would be $\mu = 1.23 \pm 0.18$.

\begin{figure}[htb!]
\center
   \includegraphics[width=.45\textwidth]{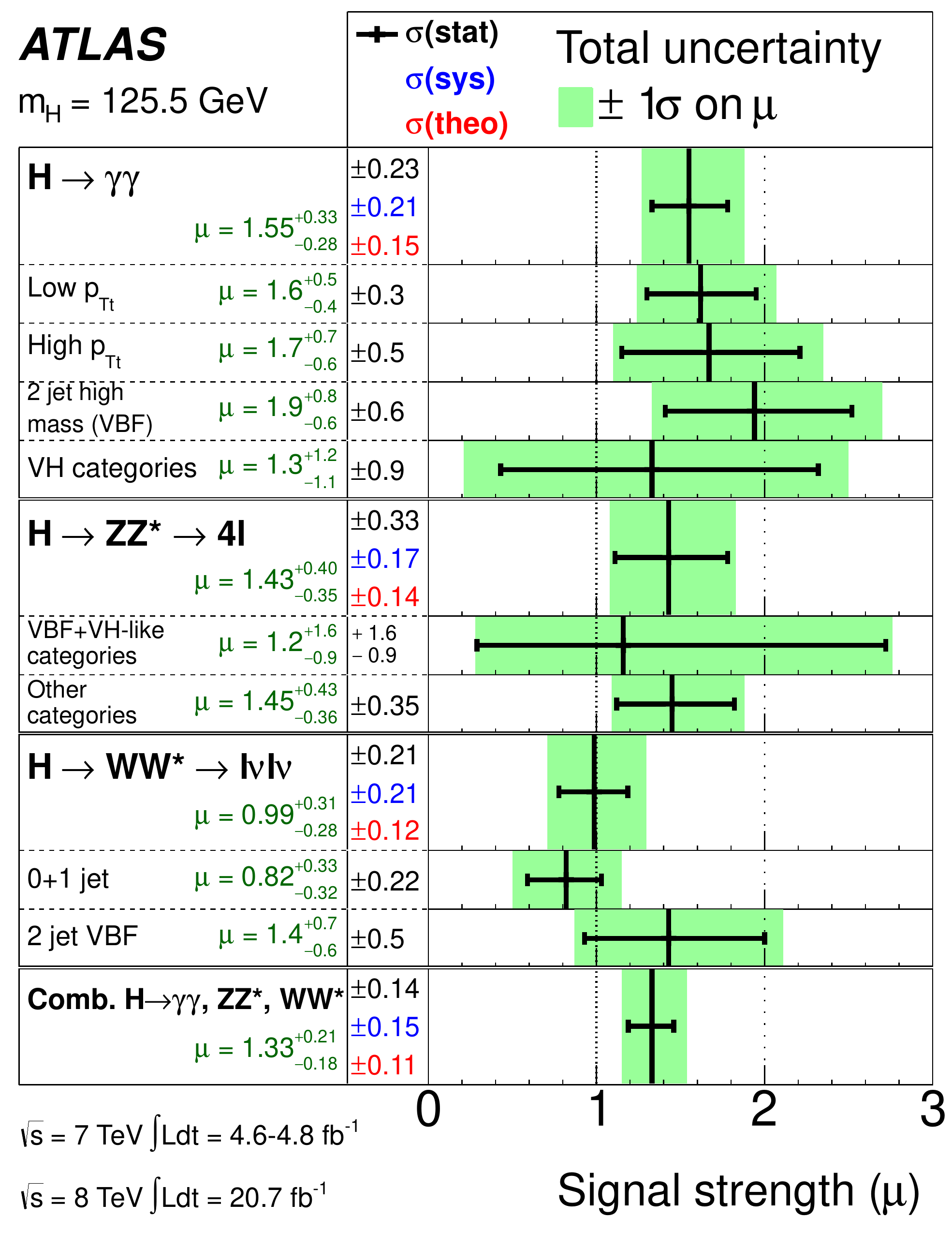}
   \caption{The measured production strengths for a Higgs boson of mass \mh\ =125.5~\GeV, normalised to 
    the SM expectations, for the individual diboson final states and their combination.  Results are also given for 
    the main categories of each analysis (described in Sections~\ref{hggcat},~\ref{sec:eventCategorisation2012} 
    and~\ref{sec:hww-njet}). The best-fit values are shown by 
     the solid vertical lines, with the total $\pm1\sigma$ uncertainty indicated by 
     the shaded band, 
     and the statistical uncertainty by the superimposed horizontal error bars. The numbers in the second
     column specify the contributions of the (symmetrised) statistical uncertainty (top), 
     the total (experimental and theoretical) systematic uncertainty (middle), and the theory uncertainty (bottom) 
     on the signal cross section (from QCD scale, PDF, and branching ratios) alone; for the individual 
     categories only the statistical uncertainty is given.}
   \label{fig:muhat_overview_WWZZgg_long}
\end{figure}

%% file: VBF.tex
The measurements of the signal strengths described in the
 previous section do not give direct information on the
relative contributions of the different production mechanisms.  Furthermore,
fixing the ratios of the production cross sections for the various
processes to the values predicted by the Standard Model may conceal 
tensions between the data and 
the theory. Therefore, in addition to the signal strengths 
for different decay modes, the
signal strengths of different production processes contributing
to the same decay mode\footnote{Such an approach avoids model assumptions 
needed for a consistent parameterisation of production 
and decay modes in terms of Higgs boson couplings.} are determined, 
exploiting the sensitivity offered
by the use of event categories in the analyses of the three channels.  

\begin{figure}[htb]
 \centering
  \includegraphics[width=.45\textwidth]{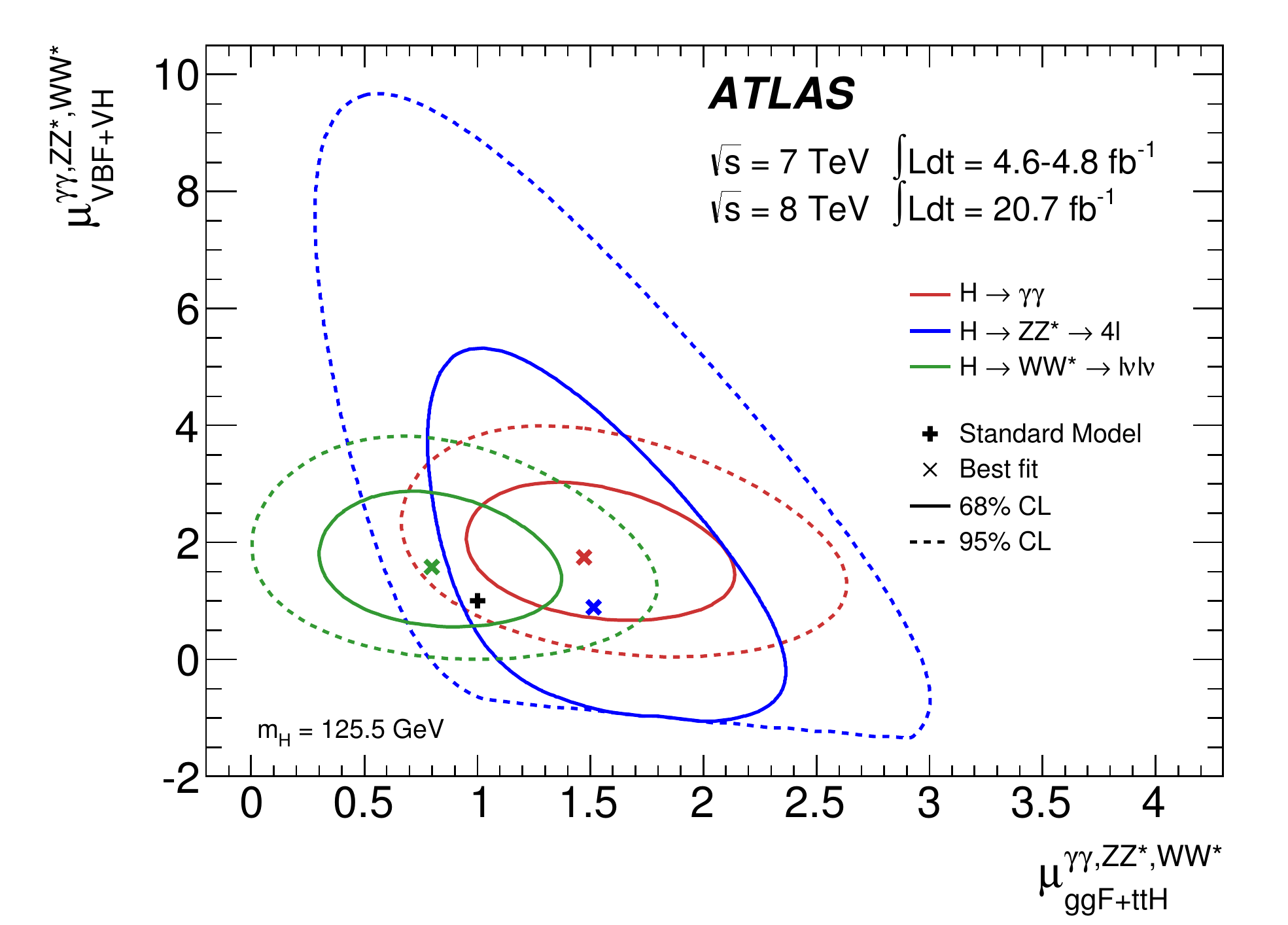}
\caption{Likelihood contours in the $(\mu^{f}_{\mathrm{ggF}+ttH},
  \mu^{f}_{\mathrm{VBF}+VH})$ plane for the final states $f$=\hgg, \hZZllll, \hWWlnln\ 
  and a Higgs boson 
  mass $m_H=125.5$~\GeV. 
The sharp lower edge of the \hZZllll\  contours
is due to the small number of events in this channel and the requirement of a positive pdf. 
   The best fits to the
  data ($\times$) and the 68\%\ (full) and 95\%\ (dashed) CL contours are
  indicated, as well as the SM expectation (+).} 
    \label{fig:ProdContourProf}
\end{figure}

\begin{figure}[htb!]
\center
   \includegraphics[width=.45\textwidth]{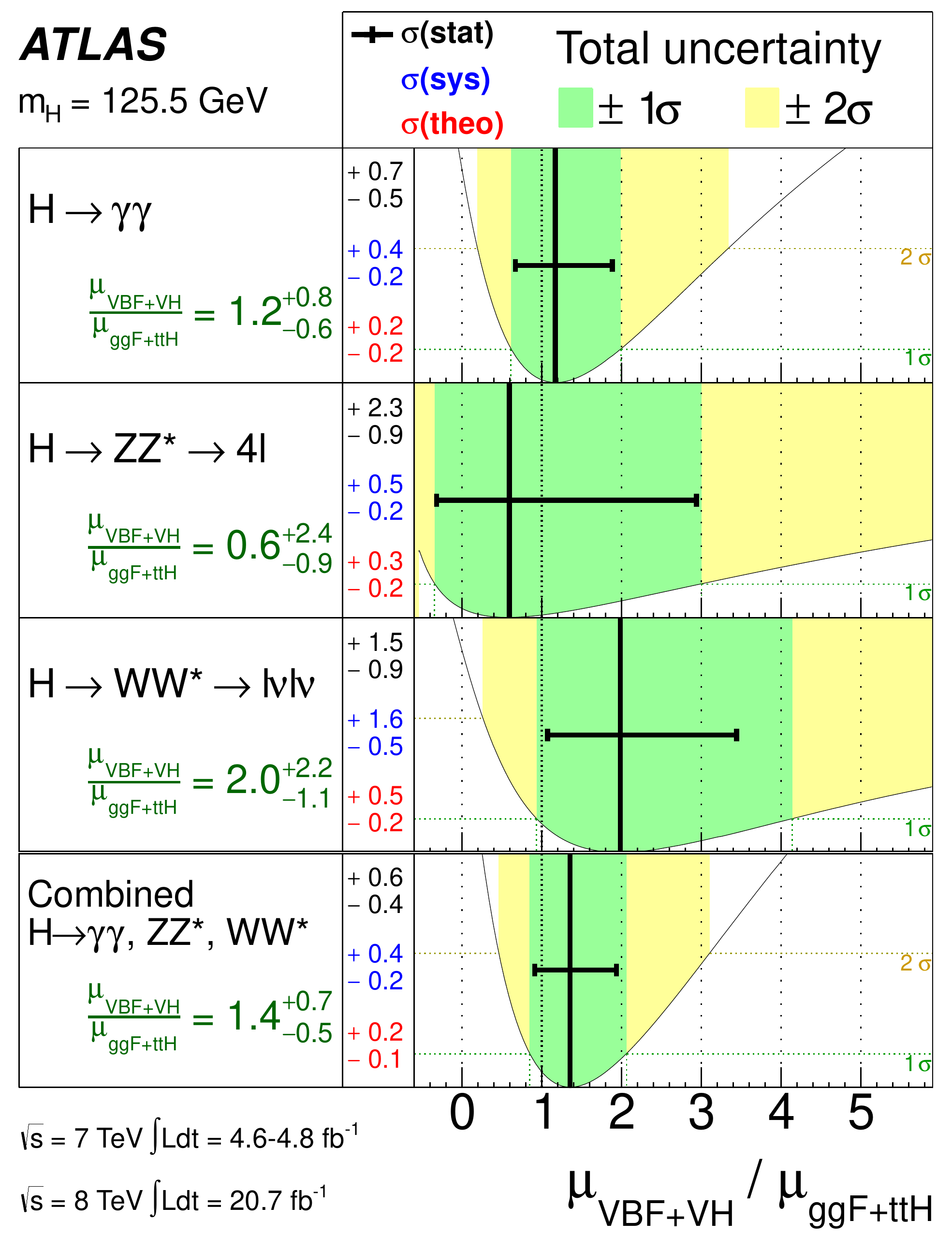}
   \caption{Measurements of the $\mu_{\mathrm{VBF}+VH}/\mu_{\mathrm{ggF}+ttH}$ ratios for  
    the individual diboson final states and their combination, for a Higgs boson mass \mh\ =125.5~\GeV. 
      The best-fit values
     are represented by the solid vertical lines, with 
     the total $\pm1\sigma$ and $\pm2\sigma$ uncertainties indicated
     by the dark- and light-shaded band, respectively, and the statistical uncertainties
     by the superimposed horizontal error bars. The numbers in the second
     column specify the contributions of the statistical uncertainty (top), 
      the total (experimental and theoretical) systematic uncertainty (middle), and the
     theoretical uncertainty (bottom) on the signal cross section (from QCD scale, PDF, 
     and branching ratios) alone.
     For a more complete illustration, 
     the distributions of the likelihood ratios 
     from which the total uncertainties are extracted are overlaid.
     }
   \label{fig:VBFoGGF_overview_short}
\end{figure}

 The data are fitted separating vector-boson-mediated processes, VBF and $VH$,
from gluon-mediated processes, ggF and $ttH$, involving fermion (mainly
top-quark) loops or legs.\footnote{Such a separation is possible 
under the assumption that the kinematic properties 
of these production modes agree with the SM predictions within uncertainties.}
Two signal strength parameters, $\mu^{f}_{\text{ggF}+ttH}=\mu^{f}_{\text{ggF}}=\mu^{f}_{ttH}$ 
and $\mu^{f}_{\text{VBF}+VH}=\mu^{f}_{\text{VBF}}=\mu^{f}_{VH}$, which scale the 
SM-predicted rates to those observed, are introduced for each of 
the considered final states ($f$=\hgg, \hZZllll, \hWWlnln). 
 The results are shown in Fig.~\ref{fig:ProdContourProf}. The 95\% CL contours
 of the measurements are consistent with the SM expectation. 
A combination of all channels would provide a higher-sensitivity test of the theory. This can
be done in a model-independent way ({\it i.e.}~without assumptions on the Higgs boson branching ratios)
by measuring the ratios 
$\mu_{\mathrm{VBF}+VH}/\mu_{\mathrm{ggF}+ttH}$ 
for the individual final states and their combination. 
The results of the fit to the data with 
the likelihood $\Lambda(\mu_{\mathrm{VBF}+VH}/\mu_{\mathrm{ggF}+ttH})$
are shown in Fig.~\ref{fig:VBFoGGF_overview_short}. Good agreement with the SM expectation 
is observed for the individual final states and their combination.

\begin{figure}[htb!]
 \centering
    \includegraphics[width=.45\textwidth]{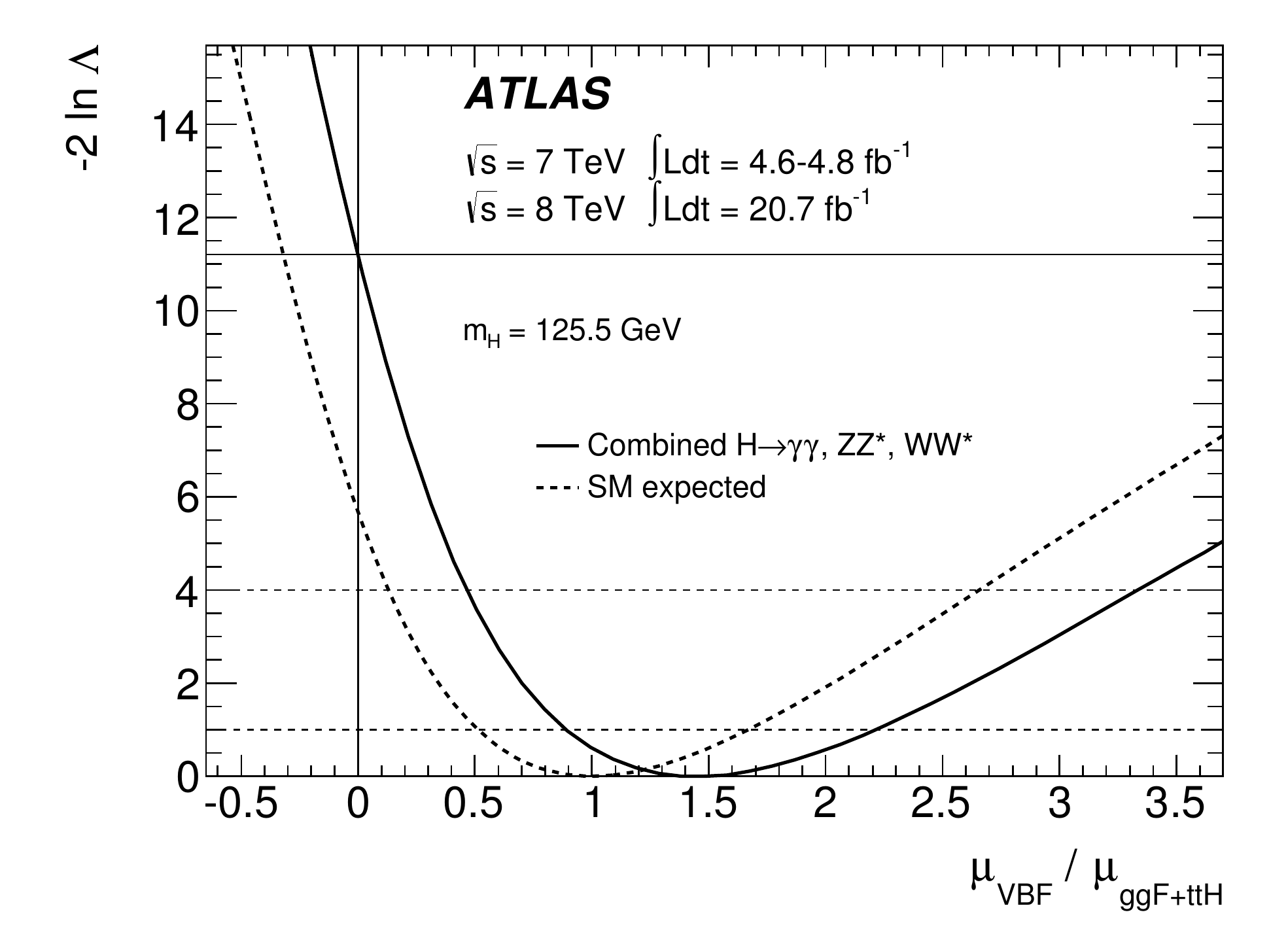} 
    \caption{Likelihood curve for the ratio $\mu_{\mathrm{VBF}}/\mu_{\mathrm{ggF}+ttH}$ for 
      the combination of the
      $H\to\gamma\gamma$, \hZZllll\ and \hWWlnln\ channels and a Higgs boson mass $m_H=125.5$~\GeV.  
      The parameter $\mu_{VH}/\mu_{\mathrm{ggF}+ttH}$ is profiled in the fit. 
      The dashed curve shows the SM expectation. 
      The horizontal dashed lines indicate the 68\% and 95\% CL.
      \label{fig:ProdRatioProf}
      }
\end{figure}

To test the sensitivity to VBF production alone, the data are also fitted with 
the ratio $\mu_{\mathrm{VBF}}/\mu_{\mathrm{ggF}+ttH}$. 
 A value 
 
\begin{equation}
\mu_{\mathrm{VBF}}/\mu_{\mathrm{ggF}+ttH}=1.4^{+0.6}_{-0.5}\,\text{(stat)}\,^{+0.5}_{-0.3}\,\text{(sys)}\label{eq:VBFratio}
\end{equation}  

\noindent is obtained from the combination of the three channels 
 (Fig.~\ref{fig:ProdRatioProf}), where the main components of the systematic 
 uncertainty come from the theoretical 
predictions for the ggF contributions to the various categories and jet multiplicities
and the knowledge of the jet energy scale and resolution.
 This result provides evidence at the  $3.3\sigma$ level that a fraction of Higgs boson 
 production occurs through VBF (as Fig.~\ref{fig:ProdRatioProf} shows, the probability for
a vanishing value of $\mu_{\mathrm{VBF}}/\mu_{\mathrm{ggF}+ttH}$, given the 
observation in the data, is 0.04\%). 
 The inclusion of preliminary $H\to\tau\tau$ results~\cite{tautauHCP}, which 
also provide some sensitivity to this ratio, would give a 
significance of $3.1\sigma$.

%% file: CouplingIntro.tex
Following the approach and benchmarks recommended in 
Refs.~\cite{Heinemeyer:2013tqa}, 
measurements of couplings are implemented using a leading-order 
tree-level motivated framework. This framework is based on the 
following assumptions:
\begin{itemize}
  \item The signals observed in the different search channels
    originate from a single resonance. A mass of $125.5\GeV$ is assumed here; the impact
    of the uncertainty reported in Eq.~(\ref{eq:mass}) on the results discussed in this
    section is negligible. 
  \item The width of the Higgs boson is narrow, 
    justifying the use of the zero-width approximation.
    Hence the predicted rate for a given channel can be decomposed in the following way:
    
    \begin{equation}
      \sigma\cdot B\ (i\to\PH\to f) = \frac{\sigma_{i}\cdot\Gamma_{f}}{\Gamma_{\PH}} 
    \end{equation}
    
    where $\sigma_{i}$ is the production cross section 
    through the initial state $i$, $B$ and $\Gamma_{f}$ are the branching ratio 
    and partial decay width into the final state $f$, respectively, and $\Gamma_{\PH}$ 
    the total width of the Higgs boson.
  \item Only modifications of coupling strengths
    are considered, while the tensor structure of the Lagrangian
    is assumed to be the same as in the Standard Model. This implies in
    particular that the observed state is a \cal{CP}-even 
    scalar.\footnote{The spin-\cal{CP} hypothesis is
     addressed in Ref.~\cite{ATLASspinpaper}.} 
\end{itemize}

The coupling scale factors $\Cc_j$ are defined in such a way that the cross sections
$\sigma_{j}$ and the partial decay widths $\Gamma_{j}$ associated with the SM particle $j$ scale
with $\Cc_j^2$ compared to the  
SM prediction~\cite{Heinemeyer:2013tqa}. With this notation,
 and with $\Cc_{\PH}^2$ being the scale factor 
 for the total Higgs boson width $\Gamma_{\PH}$,
 the cross section for the $\Pg\Pg\to\PH\to\PGg\PGg$ process, for example, can be expressed as:

\begin{eqnarray}
  \frac{\sigma\cdot B~(\Pg\Pg\to\PH\to\PGg\PGg)}{\sigma_\text{SM}(\Pg\Pg\to\PH) \cdot B_\text{SM}(\PH\to\PGg\PGg)} 
  &=&
  \frac{\Cc_{\Pg}^2 \cdot \Cc_{\PGg}^2}{\Cc_{\PH}^2}\label{eq:cross}
\end{eqnarray}

In some of the fits, $\Cc_{\PH}$ and the effective scale factors 
$\Cc_{\PGg}$ and $\Cc_{\Pg}$ for the loop-induced $\PH\to\PGg\PGg$ and $\Pg\Pg\to\PH$
 processes are expressed as a function of the more fundamental  factors 
$\Cc_{\PW}$, $\Cc_{\PZ}$, $\Cc_{\PQt}$, $\Cc_{\PQb}$ and $\Cc_{\PGt}$ (only the dominant fermion 
contributions are indicated here for simplicity). 
The relevant relationships are:

\begin{eqnarray}
 \Cc_{\Pg}^2(\Cc_{\PQb}, \Cc_{\PQt}) &=& \frac{\Cc_{\PQt}^2\cdot\sigma_{\MyggH}^{\PQt\PQt} +\Cc_{\PQb}^2\cdot\sigma_{\MyggH}^{\PQb\PQb} +\Cc_{\PQt}\Cc_{\PQb}\cdot\sigma_{\MyggH}^{\PQt\PQb}}{\sigma_{\MyggH}^{\PQt\PQt}+\sigma_{\MyggH}^{\PQb\PQb}+\sigma_{\MyggH}^{\PQt\PQb}} \label{eq:CgNLOQCD} \nonumber 
\end{eqnarray}

\begin{eqnarray}
\Cc_{\PGg}^2(\Cc_{\PQb}, \Cc_{\PQt}, \Cc_{\PGt}, \Cc_{\PW}) &=& \frac{\sum_{i,j}\Cc_i \Cc_j\cdot\Gamma_{\PGg\PGg}^{i j}}{\sum_{i,j}\Gamma_{\PGg\PGg}^{ij}} \label{eq:CgammaNLOQCD} 
\end{eqnarray}

\begin{eqnarray}
  \Cc_{\PH}^2 &=& \sum\limits_{\begin{array}{r}\scriptstyle jj=\PW\PW^{*},\ \PZ\PZ^{*},\
  \PQb\PAQb,\ \PGtm\PGtp,\\\scriptstyle\PGg\PGg,\ \PZ\PGg,\ \Pg\Pg,\ \PQt\PAQt,\ \PQc\PAQc,\ \PQs\PAQs,
  \ \PGmm\PGmp\end{array}} \frac{ \Cc_j^2 \Gamma_{jj}^\text{SM}}{\Gamma_{\PH}^\text{SM}} \label{eq:CH2_def} \nonumber
\end{eqnarray}

\noindent where $\sigma_{\MyggH}^{i j}$, $\Gamma_{\PGg\PGg}^{i j}$ and $\Gamma_{ff}^\text{SM}$ are 
obtained from 
theory~\cite{YellowReport,Heinemeyer:2013tqa}.

Results are extracted from fits to the data using the profile likelihood 
ratio $\Lambda(\vec\Cc)$, where the $\Cc_{j}$ couplings
are treated either as parameters of interest or as nuisance parameters,
depending on the measurement.

The assumptions made for the various measurements 
are summarised in Table~\ref{tab:coupling_fits} and discussed
in the next sections together with the results. 

\begin{table*}[htbp]
\centering
\caption{Summary of the coupling benchmark models discussed in this paper, where 
 $\Rr_{ij}=\Cc_i/\Cc_j$, $\Cc_{ii}=\Cc_i\Cc_i/\Cc_{\PH}$, and the functional 
 dependence assumptions are: 
 $\Cc_{V}=\Cc_{\PW}= \Cc_{\PZ}$, $\Cc_F = \Cc_{\PQt} = \Cc_{\PQb} = \Cc_{\PGt}$ (and similarly for the
 other fermions), $\Cc_{\Pg} =
 \Cc_{\Pg}(\Cc_{\PQb}, \Cc_{\PQt})$,  $\Cc_{\PGg}=\Cc_{\PGg}(\Cc_{\PQb}, \Cc_{\PQt}, \Cc_{\PGt},
 \Cc_{\PW})$, and $\Cc_{\PH}=\Cc_{\PH}(\Cc_{i})$. The tick marks indicate which assumptions are made
 in each case. The last column shows, as an example,  the relative couplings 
   involved in the $gg\to\hgg$ process, see Eq.~(\ref{eq:cross}), and
   their functional dependence in the various benchmark models.\label{tab:coupling_fits}}
\vspace{0.3cm}
\scalebox{0.93}{
\begin{tabular}{l|p{0.178\linewidth}|c|c|c|c|c|c|c}
\hline\hline
Model & Probed  & Parameters of & \multicolumn{5}{c|}{Functional assumptions} & Example: $gg\to\hgg$ \\
    &   couplings     & interest      & $\Cc_{V}$ & $\Cc_{F}$ & $\Cc_{\Pg}$ & $\Cc_{\PGg}$ & $\Cc_{\PH}$ &  \\ \hline 
1&\multirow{2}{0.99\linewidth}{Couplings to fermions and bosons} & $\Cc_{V}$, $\Cc_{F}$ & $\surd$ & $\surd$ & $\surd$ &
$\surd$ & $\surd$ & $\Cc_{F}^2 \cdot \Cc_{\PGg}^2(\Cc_{F},\Cc_{V})/ \Cc_{\PH}^2(\Cc_{F},\Cc_{V})$ \\[0.1em]\cline{1-1}\cline{3-9}
2& & $\Rr_{FV}$, $\Cc_{VV}$ & $\surd$ & $\surd$ & $\surd$ & $\surd$ & - & $\Cc_{VV}^2 \cdot \Rr_{FV}^2 \cdot \Cc_{\PGg}^2(\Rr_{FV},\Rr_{FV},\Rr_{FV},1)$ \\[0.1em]\hline
3&\multirow{2}{0.99\linewidth}{Custodial symmetry} & $\Rr_{\PW\PZ}$, $\Rr_{F\PZ}$, $\Cc_{\PZ\PZ}$ & - & $\surd$ & $\surd$ & $\surd$ & - &  $\Cc_{\PZ\PZ}^2 \cdot \Rr_{F\PZ}^2 \cdot \Cc_{\PGg}^2(\Rr_{FZ},\Rr_{FZ},\Rr_{FZ},\Rr_{\PW\PZ})$ \\[0.1em]\cline{1-1}\cline{3-9}
4& & $\Rr_{\PW\PZ}$, $\Rr_{F\PZ}$, $\Rr_{\PGg\PZ}$, $\Cc_{\PZ\PZ}$ & - & $\surd$ & $\surd$ & - & - & $\Cc_{\PZ\PZ}^2 \cdot \Rr_{F\PZ}^2 \cdot \Rr_{\PGg\PZ}^2$ \\[0.1em]\hline
5&Vertex loops & $\Cc_{\Pg}$, $\Cc_{\PGg}$ & =1 & =1 & - & - & $\surd$ & $\Cc_{\Pg}^2 \cdot \Cc_{\PGg}^2/ \Cc_{\PH}^2(\Cc_{\Pg},\Cc_{\PGg})$ \\[0.1em]\hline
%
%6&Probing BSM decays & $\Cc_{\Pg}$, $\Cc_{\PGg}$, $\BRinv$ & =1 & =1 & - & - & - & $\Cc_{\Pg}^2 \cdot \Cc_{\PGg}^2/ \Cc_{\PH}^2(\Cc_{\Pg},\Cc_{\PGg}) \cdot (1-\BRinv)$ \\[0.1em]\hline
%
%
\hline
\end{tabular}
}
\end{table*}

%% file: CVCF.tex
The first benchmark considered here (indicated as 
model~1 in Table~\ref{tab:coupling_fits}) assumes 
one coupling scale factor for fermions, $\Cc_F$, and one for bosons, 
$\Cc_V$; in this scenario, 
the \hgg\ and \ggF\ loops and the total Higgs boson width
depend only on $\Cc_F$ and $\Cc_V$, with no contributions
from physics beyond the Standard Model (BSM).  
 The strongest constraint on $\Cc_F$ comes 
indirectly from the \ggF\ production loop. 

\begin{figure}[ptb]
  \center
  \hspace*{-1cm}\includegraphics[width=.5\textwidth]{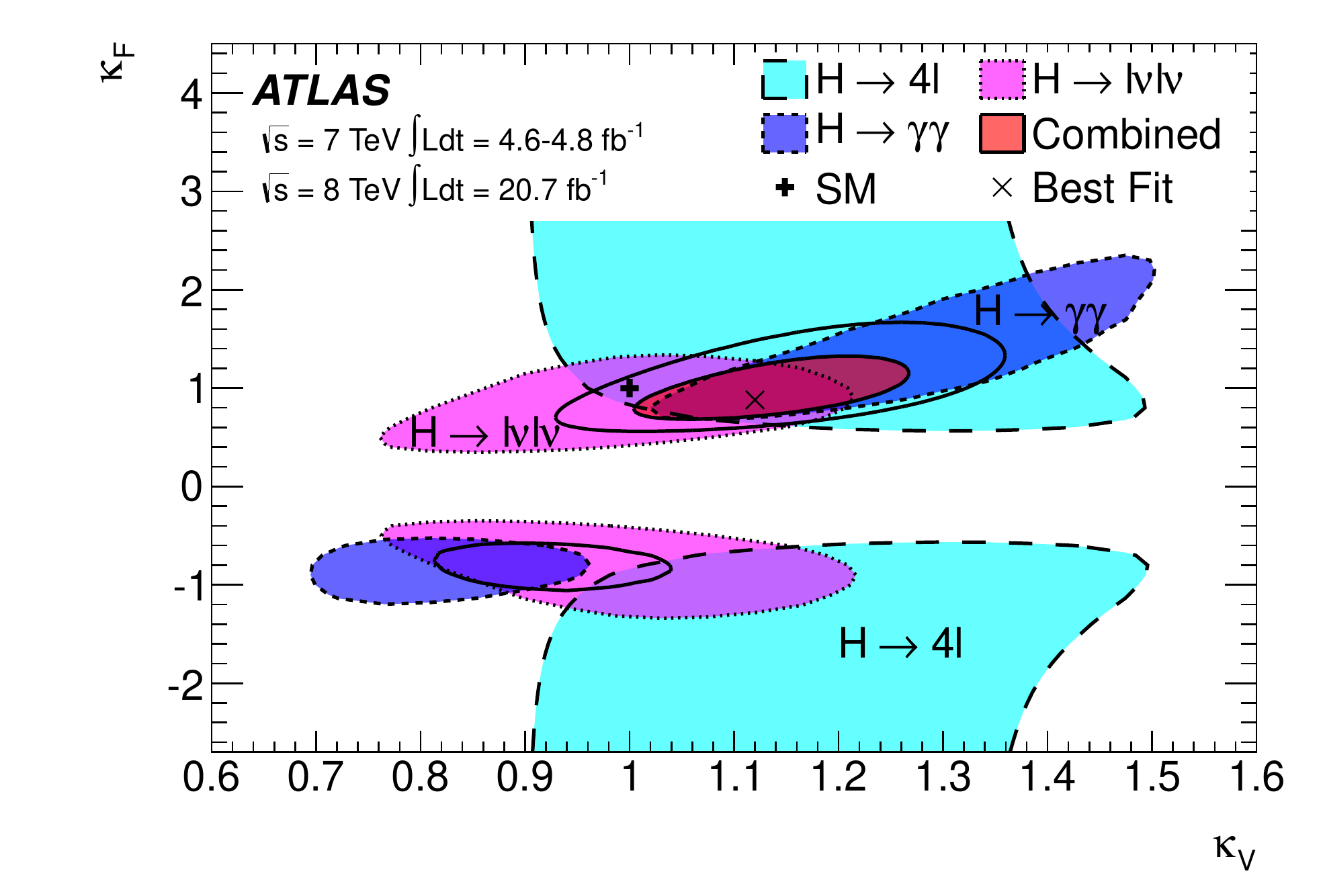}
\vspace*{-0.5cm}
  \caption{ 
    Likelihood contours (68\% CL) of the coupling scale 
    factors $\Cc_{F}$ and $\Cc_{V}$ for fermions and bosons 
    (benchmark model~1 in Table~\ref{tab:coupling_fits}), 
     as obtained from fits to the three individual channels and 
    their combination (for the latter, the 95\% CL contour is also shown). 
    The best-fit result ($\times$) and the SM expectation (+) are also indicated.
    \label{fig:spbm:CVCF}}
\end{figure}

Figure~\ref{fig:spbm:CVCF} shows the results of the fit to the
data for the three channels 
and their combination. Since only the relative sign 
of $\Cc_F$ and $\Cc_V$  is physical, in the following 
$\Cc_V>0$ is assumed. Some sensitivity to this relative sign 
is provided by the negative interference between the $W$-boson loop
and $\PQt$-quark loop in the \hgg\ decay. 
The data prefer the minimum with positive relative 
sign, which is consistent with the SM prediction, 
 but the local minimum with negative sign is 
also compatible with the observation (at the $\sim2\sigma$ level). The 
two-dimensional compatibility of the SM prediction 
with the best-fit value is 12\%.
The 68\%~CL intervals of $\Cc_F$ and $\Cc_V$, obtained by profiling over the
other parameter, are:
\begin{eqnarray}
  \Cc_F &\in& [0.76, 1.18]\\
  \Cc_V &\in& [1.05, 1.22] 
\end{eqnarray}
 with similar contributions from the statistical and systematic uncertainties. 
 
In this benchmark model, the assumption of no 
contributions from new particles to the Higgs boson width provides
strong constraints on the fermion coupling $\Cc_F$, as about 75\% 
of the total SM width comes from decays to fermions or involving
fermions.  If this assumption is relaxed, only
the ratio $\Rr_{FV}=\Cc_{F}/\Cc_{V}$ can be measured 
(benchmark model~2 in Table~\ref{tab:coupling_fits}), which 
still provides useful information on
the relationship between Yukawa and gauge couplings. 
 Fits to the data give the following 68\%\ CL intervals 
 for $\Rr_{FV}$ and $\Cc_{VV}=\Cc_{V}\Cc_{V}/\Cc_{\PH}$ (when profiling over the other parameter):
 
\begin{eqnarray}
  \Rr_{FV} &\in& [0.70, 1.01] \\
  \Cc_{VV} &\in& [1.13, 1.45] 
\end{eqnarray}

The two-dimensional compatibility of the SM prediction
with the best-fit value is 12\%. These results also exclude  
vanishing couplings of the Higgs boson to fermions (indirectly, mainly
through the \ggF\ production loop)
by more than $5\sigma$.

%% file: Custodial.tex
In the Standard Model, custodial symmetry imposes the constraint that the
$W$ and $Z$ bosons have related couplings to the Higgs boson, $g_{HVV}\sim m_V^2/{\rm v}$ 
(where v is the vacuum expectation value of the Higgs field), and that 
$\rho=m_W^2/(m_Z^2\cdot\cos^2\theta_W)$ (where $\theta_W$ is the weak Weinberg angle)
is equal to unity (as measured at LEP~\cite{EWrho}). The former
constraint is tested here by measuring the ratio
$\Rr_{\PW\PZ} =\Cc_{\PW}/\Cc_{\PZ}$. 

The simplest and most model-independent approach is to extract the ratio of 
branching ratios normalised to their SM expectation,  $\Rr_{\PW\PZ}^2=\mathrm{B}(\hww)/\mathrm{B}(\hzz)\cdot\mathrm{B}_{\rm SM}(\hzz)/\mathrm{B}_{\rm SM}(\hww)$, 
from the measured inclusive rates of the \hww\ and \hzz\ channels. 
A fit to the data with the likelihood 
$\Lambda(\Rr_{\PW\PZ})$, where $\mu_{\mathrm{ggF}+ttH}\times\mathrm{B(\hzz})/\mathrm{B}_{\rm SM} (\hzz)$
and $\mu_{\mathrm{VBF}+VH}/\mu_{\mathrm{ggF}+ttH}$ are profiled, gives
$\Rr_{\PW\PZ}= 0.81^{+0.16}_{-0.15}$.

A more sensitive measurement can be obtained by also using 
information from $WH$ and $ZH$ production, from the VBF process (which 
in the SM is roughly 
75\% $W$-fusion and 25\% $Z$-fusion mediated) and from the \hgg\ decay mode.
 A fit to the data using benchmark model~3 in Table~\ref{tab:coupling_fits}
gives the likelihood curve shown in Fig.~\ref{fig:bm:CW/CZ,CF/CZ,CZ2/CH:CW/CZ},
 with $\Rr_{\PW\PZ} \in [0.61, 1.04]$ at the 68\% CL, dominated by the statistical 
 uncertainty; the 
 other parameters, $\Rr_{F\PZ}$ and $\Cc_{\PZ\PZ}$,
are profiled. The three-dimensional compatibility of the SM prediction with 
the best-fit value is 19\%.
  \begin{figure}[htbp]
    \center
    \includegraphics[width=.45\textwidth]{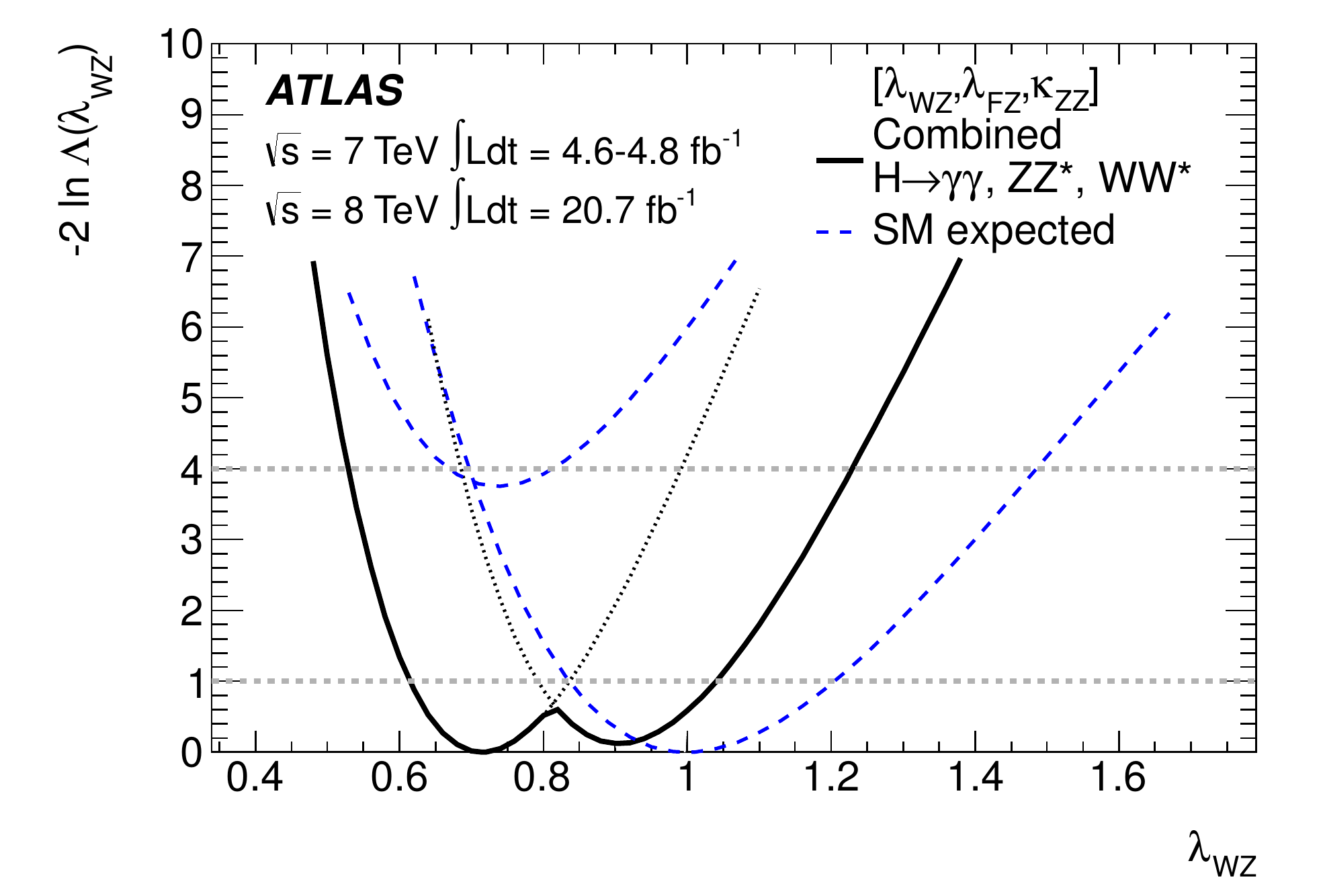}
    \vspace*{-0.5cm}
    \caption{Likelihood curve for the coupling scale factor $\Rr_{\PW\PZ}$ (benchmark 
      model~3 in Table~\ref{tab:coupling_fits}). 
      The thin dotted lines indicate the continuation of the likelihood curve 
      when restricting $\Rr_{F\PZ}$ to be either positive or negative. The dashed curves show 
      the SM expectation with the right (left) minimum indicating 
      $\Rr_{F\PZ}$ positive (negative).\label{fig:bm:CW/CZ,CF/CZ,CZ2/CH:CW/CZ}
    }
  \end{figure}

Potential contributions from BSM physics affecting
the \hgg\ channel could produce apparent deviations of the ratio $\Rr_{\PW\PZ}$ from unity
even if custodial symmetry is not broken. It is therefore 
desirable to decouple the observed \hgg\ event rate from the measurement of $\Rr_{\PW\PZ}$.
This is done with an 
extended fit for the ratio $\Rr_{\PW\PZ}$, where one extra degree of freedom 
($\Rr_{\gamma\PZ}=\Cc_{\PGg}/\Cc_{\PZ}$) absorbs 
possible BSM effects in the \hgg\ channel (benchmark model~4 in Table~\ref{tab:coupling_fits}). 
This measurement yields: 

\begin{equation}
\Rr_{\PW\PZ} = 0.82 \pm 0.15\label{eq:custodial}
\end{equation}  

\noindent and a four-dimensional compatibility of the SM prediction with the best-fit value of 20\%.

%% file: KgammaKg.tex
Many BSM physics scenarios predict the existence of new heavy particles, which can
contribute to loop-induced processes such as \ggF\ production and \hgg\ decay. 
 In the approach used here (benchmark model~5 in Table~\ref{tab:coupling_fits}), 
  it is assumed that the new particles do not contribute to the Higgs boson width and that the  
 couplings of the known particles to the Higgs boson have SM strength 
 ({\it i.e.}~$\Cc_{i}$=1). 
  Effective scale factors 
 $\Cc_{\Pg}$ and $\Cc_{\PGg}$ are introduced to parameterise the \ggF\ and \hgg\ loops. 
  The results of their measurements from a fit to the data are shown in 
  Fig.~\ref{fig:bm:Cg,Cgamma:Cgamma,Cg}. The best-fit values 
 when profiling over the other parameters are:
 
\begin{eqnarray}
  \Cc_{\Pg} &=& 1.04 \pm 0.14\\
  \Cc_{\PGg} &=& 1.20 \pm 0.15
\end{eqnarray}

The two-dimensional compatibility of the SM prediction with the best-fit value
is 14\%.

  \begin{figure}[htbp!]
    \center
      \includegraphics[width=.45\textwidth]{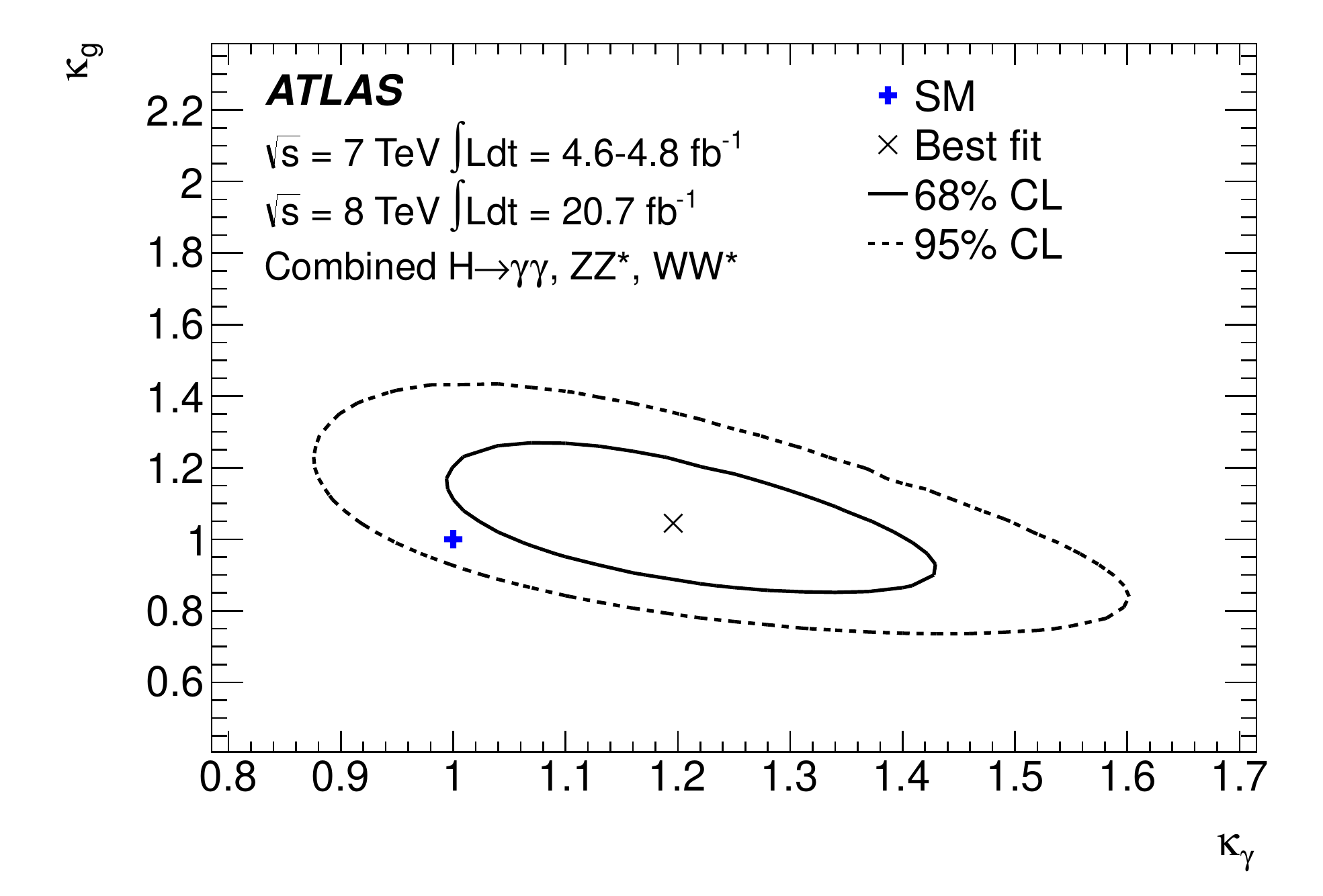}
      \vspace*{-0.5cm}
    \caption{Likelihood contours for the coupling scale factors $\Cc_{\PGg}$ and $\Cc_{\Pg}$ 
      probing BSM contributions to the \hgg\ and \ggF\ loops, assuming no BSM
      contributions to the total Higgs boson width (benchmark model~5 
      in Table~\ref{tab:coupling_fits}). The best-fit result~($\times$) and the SM expectation~(+) 
      are also indicated.
      \label{fig:bm:Cg,Cgamma:Cgamma,Cg}
    }
  \end{figure}

%% file: Results.tex
\begin{figure}[htb!]
   \center
   \includegraphics[width=.45\textwidth]{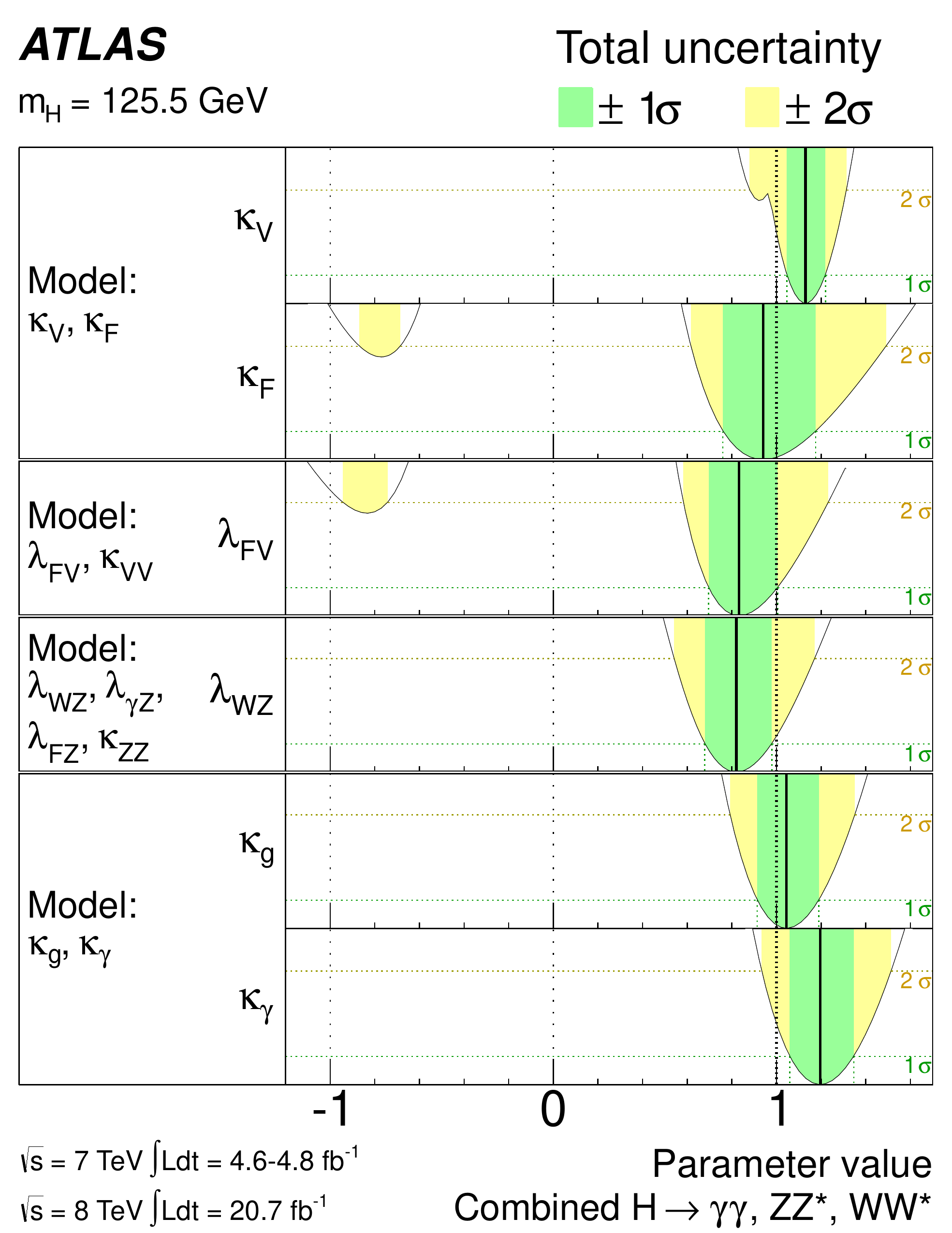}
   \caption{Summary of the measurements of the coupling scale factors for a Higgs boson with
    mass \mh=125.5~\GeV. 
            The best-fit values are represented by the solid vertical lines, 
            with the $\pm1\sigma$ and $\pm2\sigma$ uncertainties given 
            by the dark- and light-shaded band, respectively. For a more complete illustration, 
            the distributions of the likelihood ratios
            from which the total uncertainties are extracted are overlaid.
            The measurements in the various benchmark models, 
	    separated by double horizontal lines, 
            are strongly correlated.
   }
   \label{fig:coupling_summary}
\end{figure}

The results of the measurements of the coupling scale factors 
discussed in the previous sections, obtained
 under the assumptions detailed in 
 Section~\ref{Couplings} and Table~\ref{tab:coupling_fits},  
 are summmarised in Fig.~\ref{fig:coupling_summary}. The measurements in the various benchmark 
models are strongly correlated, as they are obtained from fits to the same experimental data. 
 A simple $\chi^2$-like compatibility test with the SM is therefore 
 not meaningful.

  The coupling of the new particle to gauge bosons $\Cc_V$ is constrained 
  by several channels, directly and indirectly, at the $\pm$10\% level. 
 Couplings to fermions with a significance larger than $5\sigma$  are
indirectly observed mainly through the gluon-fusion production process, assuming
the loop is dominated by fermion exchange. 
 The ratio of the relative couplings of the Higgs boson to 
 the $W$ and $Z$ bosons, $\Cc_{\PW}/\Cc_{\PZ}$, is measured to be consistent
 with unity, as predicted by custodial symmetry.
 Under the hypothesis that all couplings of the Higgs boson to the 
known particles are fixed to their SM values, and assuming no 
BSM contributions to the Higgs boson width, 
  no significant anomalous
contributions to the \ggF\ and \hgg\ loops are observed.

%% file: acknowledgements.tex
% Acknowledgements for papers with collision data
% Version 1-Apr-2013
\section{Acknowledgements}

% Standard acknowledgements start here
%----------------------------------------------
We thank CERN for the very successful operation of the LHC, as well as the
support staff from our institutions without whom ATLAS could not be
operated efficiently.

We acknowledge the support of ANPCyT, Argentina; YerPhI, Armenia; ARC,
Australia; BMWF and FWF, Austria; ANAS, Azerbaijan; SSTC, Belarus; CNPq and FAPESP,
Brazil; NSERC, NRC and CFI, Canada; CERN; CONICYT, Chile; CAS, MOST and NSFC,
China; COLCIENCIAS, Colombia; MSMT CR, MPO CR and VSC CR, Czech Republic;
DNRF, DNSRC and Lundbeck Foundation, Denmark; EPLANET, ERC and NSRF, European Union;
IN2P3-CNRS, CEA-DSM/IRFU, France; GNSF, Georgia; BMBF, DFG, HGF, MPG and AvH
Foundation, Germany; GSRT and NSRF, Greece; ISF, MINERVA, GIF, DIP and Benoziyo Center,
Israel; INFN, Italy; MEXT and JSPS, Japan; CNRST, Morocco; FOM and NWO,
Netherlands; BRF and RCN, Norway; MNiSW, Poland; GRICES and FCT, Portugal; MERYS
(MECTS), Romania; MES of Russia and ROSATOM, Russian Federation; JINR; MSTD,
Serbia; MSSR, Slovakia; ARRS and MIZ\v{S}, Slovenia; DST/NRF, South Africa;
MICINN, Spain; SRC and Wallenberg Foundation, Sweden; SER, SNSF and Cantons of
Bern and Geneva, Switzerland; NSC, Taiwan; TAEK, Turkey; STFC, the Royal
Society and Leverhulme Trust, United Kingdom; DOE and NSF, United States of
America.

The crucial computing support from all WLCG partners is acknowledged
gratefully, in particular from CERN and the ATLAS Tier-1 facilities at
TRIUMF (Canada), NDGF (Denmark, Norway, Sweden), CC-IN2P3 (France),
KIT/GridKA (Germany), INFN-CNAF (Italy), NL-T1 (Netherlands), PIC (Spain),
ASGC (Taiwan), RAL (UK) and BNL (USA) and in the Tier-2 facilities
worldwide.
%----------------------------------------------

%% file: atlas_authlist.tex
% ATLAS Collaboration author list
% Data extracted on 14-Oct-2014 for paper reference HIGG-2013-02
%\documentclass[11pt]{article}
%\usepackage{a4wide}\begin{document}
\begin{flushleft}
{\Large The ATLAS Collaboration}

\bigskip

G.~Aad$^{\rm 48}$,
T.~Abajyan$^{\rm 21}$,
B.~Abbott$^{\rm 112}$,
J.~Abdallah$^{\rm 12}$,
S.~Abdel~Khalek$^{\rm 116}$,
O.~Abdinov$^{\rm 11}$,
R.~Aben$^{\rm 106}$,
B.~Abi$^{\rm 113}$,
M.~Abolins$^{\rm 89}$,
O.S.~AbouZeid$^{\rm 159}$
% ...
\bigskip
\\
% ...
$^{11}$ Institute of Physics, Azerbaijan Academy of Sciences, Baku, Azerbaijan\\
$^{12}$ Institut de F{\'\i}sica d'Altes Energies and Departament de F{\'\i}sica de la Universitat Aut{\`o}noma de Barcelona, Barcelona, Spain\\
% ...
$^{21}$ Physikalisches Institut, University of Bonn, Bonn, Germany\\
% ...
$^{48}$ Fakult{\"a}t f{\"u}r Mathematik und Physik, Albert-Ludwigs-Universit{\"a}t, Freiburg, Germany\\
% ...
$^{89}$ Department of Physics and Astronomy, Michigan State University, East Lansing MI, United States of America\\
% ...
$^{106}$ Nikhef National Institute for Subatomic Physics and University of Amsterdam, Amsterdam, Netherlands\\
% ...
$^{112}$ Homer L. Dodge Department of Physics and Astronomy, University of Oklahoma, Norman OK, United States of America\\
$^{113}$ Department of Physics, Oklahoma State University, Stillwater OK, United States of America\\
% ...
$^{116}$ LAL, Universit{\'e} Paris-Sud and CNRS/IN2P3, Orsay, France\\
% ...
$^{159}$ Department of Physics, University of Toronto, Toronto ON, Canada\\
% ...
\end{flushleft}

%\end{document}
% Created with ./xml2latex.py